# Topological metasurface: From passive toward active and beyond

Jian Wei You,[1, *] Zhihao Lan,[2] Qian Ma,[1] Zhen Gao,[3] Yihao Yang,[4] Fei Gao,[4] Meng Xiao,[5] and Tie Jun Cui[1, *]

[1]*State Key Laboratory of Millimetre Waves, School of Information Science and Engineering, Southeast University, Nanjing 210096, China*

[2]*Department of Electronic and Electrical Engineering, University College London, London WC1E 7JE, United Kingdom*

[3]*Department of Electronic and Electrical Engineering, Southern University of Science and Technology, Shenzhen 518055, China*

[4]*State Key Laboratory of Modern Optical Instrumentation, College of Information Science and Electronic Engineering, Zhejiang University, Hangzhou 310058, China*

[5]*School of Physics and Technology, Wuhan University, Wuhan 430072, China*

*Corresponding author: jwyou@seu.edu.cn and tjcui@seu.edu.cn*

## Abstract

Metasurfaces are subwavelength structured thin films consisting of arrays of units that allow the controls of polarization, phase and amplitude of light over a subwavelength thickness. The recent developments in topological photonics have greatly broadened the horizon in designing the metasurfaces for novel functional applications. In this review, we summarize recent progress in the research field of topological metasurfaces, firstly from the perspectives of passive and active in the classical regime, and then in the quantum regime. More specifically, we begin by examining the passive topological phenomena in two-dimensional photonic systems, including both time-reversal broken systems and time-reversal preserved systems. Subsequently, we move to discuss the cutting-edge studies of the active topological metasurfaces, including nonlinear topological metasurfaces and reconfigurable topological metasurfaces. After overviewing the topological metasurfaces in the classical regime, we show how the topological metasurfaces could provide a new platform for quantum information and quantum many-body physics. Finally, we conclude and describe some challenges and future directions of this fast-evolving field.

## 1. Introduction

Motivated by their planar architecture and great potentials for the future on-chip applications, metasurfaces have attracted great attention in recent years [1]-[9]. In essence, metasurfaces are sub-wavelength structured thin layers consisting of arrays of scatterers that can be used to control the polarization, phase and amplitude of light over a sub-wavelength thickness. In such 2D ultrathin metasurfaces, by spatially varying the geometric parameters of the scatterers, such as their shape, size, orientation, abrupt and controllable changes of optical properties can be achieved via engineering the resonant interaction between light and the scatterers, which offers a fundamentally new method of light manipulation beyond the conventional methods based on refraction or propagation in bulk materials. The metasurfaces not only provide more compact platforms for the study of light-matter interaction in reduced dimensions, but also allow for a plethora of practical applications, such as, efficient wavefront shaping, beam steering polarization control, as well as enhancement of the emission and detection of light.

On the other hand, the concept of topology has also been introduced into the realm of photonics in the past decades, giving birth to the field of topological photonics [10]-[16]. Topological photonics mainly work according to the bulk-edge correspondence principle, where nontrivial bulk properties of the photonic systems manifest through the emergence of topological edge states propagating along the system edges. These edge states are robust against backscattering due to disorder or defects, thus opening a door to design photonic devices with unprecedented performance. While the developments in topological photonics are largely motivated by the relevant topological physics and phenomena discovered in condensed matter physics, which mostly are electronic materials, due to the bosonic nature of photons, topological photonics has its own unique physics and phenomena not present in condensed matter systems.

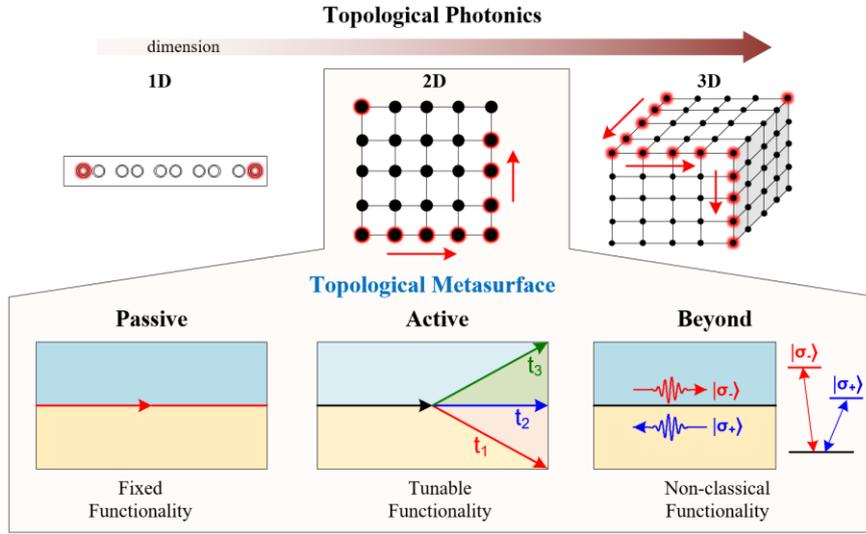

**Fig. 1.** Topological metasurface evolved from passive toward active and beyond.

Topological photonics have been extensively studied in 1D, 2D and 3D (see Fig.1 for the schematic). In this Review, we will mainly focus on the 2D topological photonic systems to illustrate how the research of metasurfaces could be greatly enriched by the integration of optical metasurfaces with topological concepts. We would like to note that though the topological physics and phenomena are much richer in 3D photonic systems, the challenges in fabricating complex nanoscale 3D photonic structures limit their practical compact applications when compared to flat optical components based on metasurfaces, which are much more compatible with the state-of-the-art planar nanofabrication processes, such as integrated photonic circuits. The field of 2D topological metasurfaces (TM) could be described from the perspectives of passive to active in the classical regime and then to the quantum regime. Here, we remark that the traditional metasurfaces refer to the two-dimensional counterparts of the conventional metamaterials, which are different from photonic crystals [17]. However, as the rapid progress in the development of metamaterials and metaphotonics, the concept of metasurface now has already been widely generalized. Currently, the generalized metasurface usually refers to a layer of artificial materials with sub-wavelength thickness, which can be either structured or unstructured with a series of patterns in the horizontal dimensions [18]-[21].

For the passive topological metasurfaces, which mainly focus on a fixed functionality, we will discuss analog quantum Hall TM, analog quantum spin Hall TM, analog quantum valley Hall TM, Floquet TM, high-order TM, and other TM, such as bound states in the continuum (BICs), Skyrmions. While passive topological metasurfaces can provide a static and predefined functionality which is fixed by the structure of the device, dynamic performance of the metasurface with tunability and reconfigurability or even the integration of multiple optical functionalities into compact metasurface platforms is preferred in most practical applications. We will give an overview on the various methods that have been used in the literature to achieve active control of the metasurface properties, such as, by mechanical, thermal, electrical, or optical methods and show how the performance of the passive topological metasurfaces could be dynamically changed. Moving to the quantum regime, we also discuss the non-classical functionalities that topological metasurface could perform, such as quantum information applications, where superposition, entanglement, interference and correlation between multi-photon state of light are crucial. By coupling quantum emitters to topological metasurfaces, not only high-efficiency quantum photon sources could be obtained, but also via manipulating many-body cooperative interactions among the emitters, strong interacting quantum many-body topological phases of light could be explored.

The review is structured as follows. In section 2, we discuss passive topological metasurfaces, including analog quantum Hall (QH) TM, analog quantum spin Hall (QSH) TM, analog quantum valley Hall (QVH) TM, Floquet TM, high-order TM, and other topological metasurfaces, such as BICs, Skyrmions. In section 3, different methods that could be used to actively tune the properties of the topological metasrufaces, such as electrical, optical, mechanical, thermal and others are described. In section 4, topological metasurfaces for applications ranging from quantum information process to metasurface-emitter coupling to quantum many-body physics of light are discussed. We conclude in section 5 and give some future prospects about this fast-evolving field of topological metasurfaces.

## 2. Passive topological metasurface

The topological metasurface is a classical-electromagnetics analog counterpart of the two-dimensional electronic topological insulators studied in condensed matter physics. As an important part of topological photonic insulators, topological metasurface can be subdivided into time-reversal (TR)-broken and TR-preserved systems. In this section, we review the passive topological metasurfaces, which mainly focus on a fixed functionality and include analog QH topological metasurfaces, QSH topological metasurfaces, QVH topological metasurfaces, Floquet topological metasurfaces, high-order topological metasurfaces, and other topological metasurfaces.

### 2.1 QH topological metasurface

An important insight into the topology is that the topological phases and topological phase transitions are not only restricted to fermions such as electrons in quantum system, but also to bosons such as photons in classical-wave systems. A prominent example was proposed by Haldane and Raghu in 2008: the robust chiral edge state of electrons in the quantum Hall effect (QHE) can also be realized in gyromagnetic photonic crystals placed in strong external magnetic field to break the time-reversal symmetry that support "one-way waveguide" allowing electromagnetic waves flow in one direction only [22]. As shown in Fig. 2(a), in a hexagonal two-dimensional (2D) array of cylindrical dielectric ferrite rods (such as yttrium iron garnet), when the external magnetic field is zero and the time-reversal symmetry is preserved, the hexagonal crystal symmetry guarantees the existence of a pair of Dirac points in the corner of 2D Brillion zone (BZ). When place the gyromagnetic photonic crystals in a strong external magnetic field to break the time-reversal symmetry, the degeneracy of the Dirac points will be lifted and each nondegenerate band will exhibit a nonzero Chern number. The nontrivial photonic bandgap will support robust "one-way" photonic chiral edge states that propagate along only one direction and are immune to sharp corners, defects and disorders. More interestingly, the total number of chiral edge states is exactly equal to the summation of the Chern numbers of photonic bands bellow the photonic bandgap, which is the famous bulk-boundary correspondence.

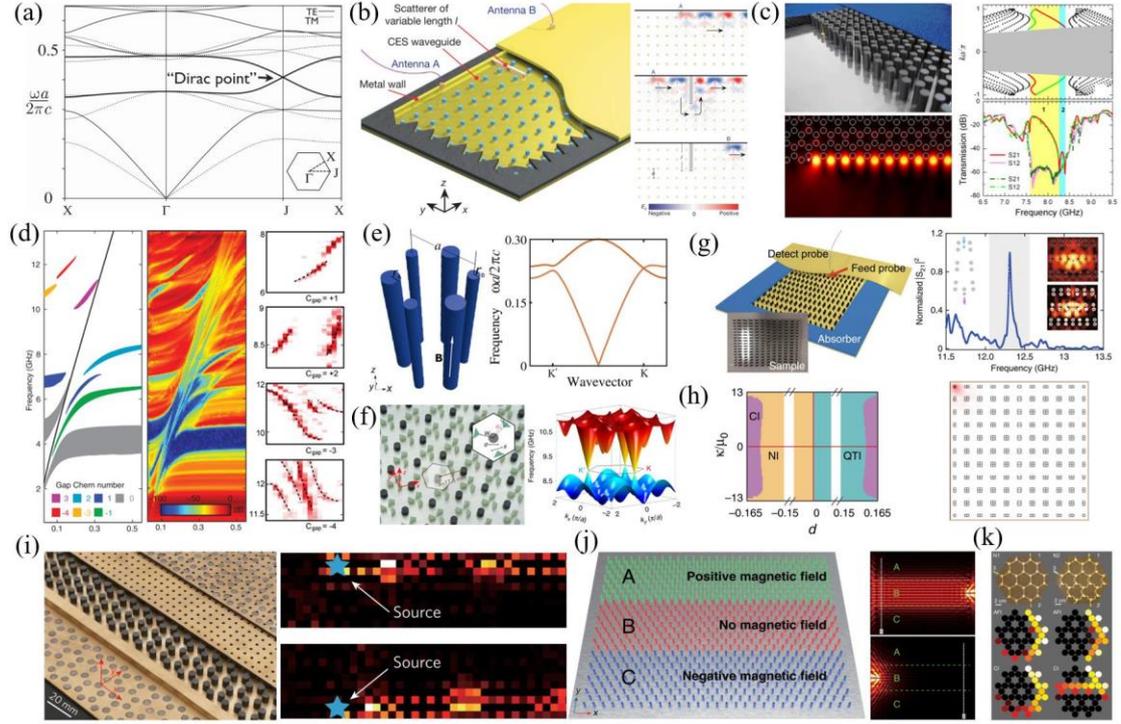

**Fig. 2.** Analog quantum Hall topological metasurface: (a) Bulk dispersion of a gyromagnetic photonic crystal under zero external magnetic field [22]. (b) Experimental set-up for measuring the one-way chiral edge state in a gyromagnetic topological metasurface [23]. (c) Gyromagnetic topological metasurface supporting self-guiding unidirectional electromagnetic edge states [24]. (d) Topological bandgap map as a function of magnetic field strength and frequency [25]. (e) Magnetic topological metasurface composed of ferromagnetic rods arranged in a honeycomb lattice [26]. (f) Band structure of magnetic topological metasurface with an unpaired Dirac point [27]. (g) Spectrum and field profiles of the dislocation-induced topological metasurface [28]. (h) Topological phase transition diagram and simulated mode profile of quadrupole topological corner state [29]. (i) Experimental set-up for measuring the antichiral edge state [30]. (j) Nonreciprocal large-area topological metasurface [31]. (k) Anomalous nonreciprocal topological metasurface made of ferrite circulators connected with microstrip lines [33].

This pioneering work gave birth to the research area of topological photonic which have revolutionized the whole field of photonics. In 2009, this fascinating idea was experimentally verified by Wang *et al.* [23] in microwave frequency for the first time by implementing a 2D magneto-optical photonic crystal consisting of a square array of ferrite rods placed in an external magnetic field, as shown in the left pane of Fig. 2(b), which will induce gyromagnetic anisotropy in the ferrite that breaks the time-reversal symmetry and support single chiral edge state in the photonic bandgap. Such chiral edge states only have positive or negative group velocities which depend on the direction of external magnetic field, implying that electromagnetic waves can only propagate in one direction without reflection and scattering from obstacles, as shown in right panel of Fig. 2(b). In this pioneering experimental work, an ancillary cladding layer made of either perfect metal or photonic crystal is needed to confine the photonic chiral edge states. As pointed out by Wang et al. [23], without the ancillary cladding, the photonic chiral edge states will disappear because of the leakage into the surrounding air. To overcome this limitation, Poo *et al.* proposed a honeycomb magnetic photonic crystal that supports self-guiding unidirectional edge states along the zigzag edge [24], as shown in left panel of Fig.2 (c). This self-guiding mechanism originates from the fact that the chiral edge states exist bellow the light cone, as shown in the right panel of Fig. 2(c), therefore, the self-guiding one-way photonic edge states are evanescent in free space and tightly confined on the zigzag edge without ancillary cladding.

Despite great progresses have been made in two-dimensional photonic Chern insulator, all previous theoretical and experimental studies have been limited to Chern number of one, which means that the photonic Chern insulators support only one chiral edge state according to the bulk-boundary correspondence. It is highly desirable to construct photonic Chern insulators with large Chern numbers and multiple chiral edge states. By simultaneously gapping multiple sets of Dirac and quadratic degeneracies, Skirlo *et al.* experimentally verified that large Chen number can be achieved in 2D magnetic photonic crystals [25], as illustrated in Fig. 2(d). According to bulk-boundary correspondence, the total number of edge states is equal to the gap Chern number ($C_{gap}$) and the sign of $C_{gap}$ is consistent with the group velocity of the chiral edge states. By Fourier transforming the mode profiles of the chiral edge states, the measured dispersions of chiral edge states shown in the right panel of Fig. 2(d) demonstrate larger gap Chern number from 2 to -4. The photonic bandgaps opened by breaking parity inversion symmetry (*P*) or time-reversal symmetry (*T*) individually are topologically inequivalent since in these two cases their bulk bands carry different topological invariants. When both *P* and *T* symmetries are broken simultaneously in a gyromagnetic photonic crystal, the competition between two broken symmetries will gives rise to fantastic physical properties such as one-way Klein tunneling and unpaired Dirac point [26]. As shown in Fig. 2 (e), the competition between broken *P* and *T* symmetries closes the gap at K' point but open the one at K point, revealing the presence of an unpaired Dirac point [26]. This theoretical prediction was experimentally verified later by Liu *et al.* [27], in which each unit cell consists of a gyromagnetic ferrite rod surrounded by three right-triangular dielectric pillars to break *T* and *P* symmetries simultaneously. Rotating three right-triangular dielectric columns will break the *P* symmetry and the external magnetic field will break the *T* symmetry, and consequently, an unpaired Dirac point appears at K' point when *P* symmetry broken is equal to *T* symmetry broken, corresponding to orientation angle of $\theta = 12.9°$ and 0.4 Tesla external magnetic field, as shown in Fig. 2(f).

Beside the robust chiral topological edge states, another type of fundamental component in photonic chips, robust topological cavity modes can also be realized in a gyromagnetic photonic crystal. Li *et al.* introduces a dislocation in a rectangular-lattice magnetic photonic crystal, and consequently, this topological cavity mode is protected by dual-topology: concurrent wavevector space and real-space topology, as shown in Fig. 2(g). This topological cavity mode was experimentally observed directly by measuring its response spectrum and field distributions, as illustrated in right panel of Fig.2(g) [28]. Beyond the dislocation-induced topological cavity modes, recently discovered higher-order topological insulators (HOTI) provide another novel ground for designing robust cavity modes. He *et al.* theoretically proposed a quadrupole-HOTI with localized zero-dimensional (0D) corner states in square-lattice magnetic photonic crystal protected by simultaneous presence of crystalline symmetries and broken time-reversal symmetry [29]. Due to the symmetry competition, the gyromagnetic photonic crystal exhibits multiple topological phases from Chern insulator to quadrupole HOTI, and the evolution of topological phase transition is shown in left panel of Fig. 2(h). The quadrupole HOTI phase supports 0D topological corner state, whose field distribution is shown in the right panel of Fig. 2(h).

From the celebrated Haldan model that exhibits the quantum Hall effect with broken time-reversal symmetry, two chiral edge states will propagate in opposite directions along two parallel edges of a magnetic photonic crystal stripe sample. However, in a modified Haldane model a counterintuitive physical phenomenon is experimentally observed that two antichiral edge states can propagate in the same direction along two parallel strip edges [30]. The underlying physical mechanism of antichiral edge states can be described by a modified Haldane model, where the next-nearest-neighbor couplings for different sublattices have opposite signs in phase. The experimental sample shown in top panel of Fig. 2(i) is a honeycomb magnetic photonic crystal with two set of sublattices magnetically biased in opposite directions. The measured antichiral edge states will propagate in the right direction along both the upper and lower stripe edges, as shown in bottom panel of Fig. 2(i). While the previously mentioned chiral or antichiral edge states propagate along a certain narrow edge, Wang *et al.* theoretically proposed and experimentally demonstrated a topological one-way large-area waveguide by sandwiching a trivial nonmagnetic photonic crystal between two nontrivial magnetic photonic crystals, as illustrated in top panel of Fig. 2(j) [31]. In contrast with conventional chiral edge state, the large-area unidirectional waveguide supports one-way waveguide states with uniformly distributed amplitude over a large area, as shown in bottom panel of Fig.2(j). Additionally, based on the coupled dipole method [32], a QH domain wall between positive and negative gyrotropy has been constructed to study the photonic analogy of Jackiw-Rebbi waves. As a counterpart of the tight-binding approximation, the coupled dipole method offers an intuitive and instructive modeling tool to study the quantum Hall phase in the optical regime.

Though the photonic Chern insulators are regarded as the most robust topological phases reported so far, most recently, Zhang *et al.* theoretically and experimentally demonstrated an anomalous non-reciprocal topological phase whose topological protection is even stronger than that of the photonic Chern insulator [33]. This novel topological phase was first theoretically identified by unitary scattering networks and then experimentally confirmed by ferrite circulators and microstrip lines network, as shown in top panel of Fig. 2(k). The superior robustness of the anomalous non-reciprocal topological edge states is directly observed by measuring its field profiles, as shown in the middle panel of Fig. 2(k), where the anomalous topological edge state is robust to disruption (phase delay abruptly changes from $\pi/8$ to $\pi/2$) and keeps transmitting to port 2, but the chiral edge state of photonic Chern insulator changes propagation direction and travels along the interface to port 3 in the presence of the disruption, as presented in bottom panel of Fig. 2(k).

## 2.2 QSH topological metasurface

The conception of quantum spin Hall effect (QSHE) was first proposed in condensed matters [34] and observed in semiconductor[35][36], which can be regarded as two copies of quantum Hall states, where electrons feel spin-dependent magnetic fields. In a quantum spin Hall (QSH) system, time reversal symmetry is preserved, which releases the requirement of strong magnetic fields in quantum Hall systems. Thanks to the spin-1/2 of electrons, time reversal symmetry $T_f^2 = -1$ guarantees the Kramers doublet of electronic states composed of spin-up and spin-down states at the time-reversal-invariant points. When SOC is considered in the QSH system, spin-dependent gauge fields are introduced to unfold the Kramers double degeneracy of electronic states away from the time-reversal-points, while the degeneracy at the time-reversal-invariant-points is still degenerate. For a finite-size system composed of two regions with different topological invariants, two topological edge states with opposite spins and momenta can be supported, each of which is spin-momentum locked, backscattering immune, and robust against the perturbations that avoiding spin flipping.

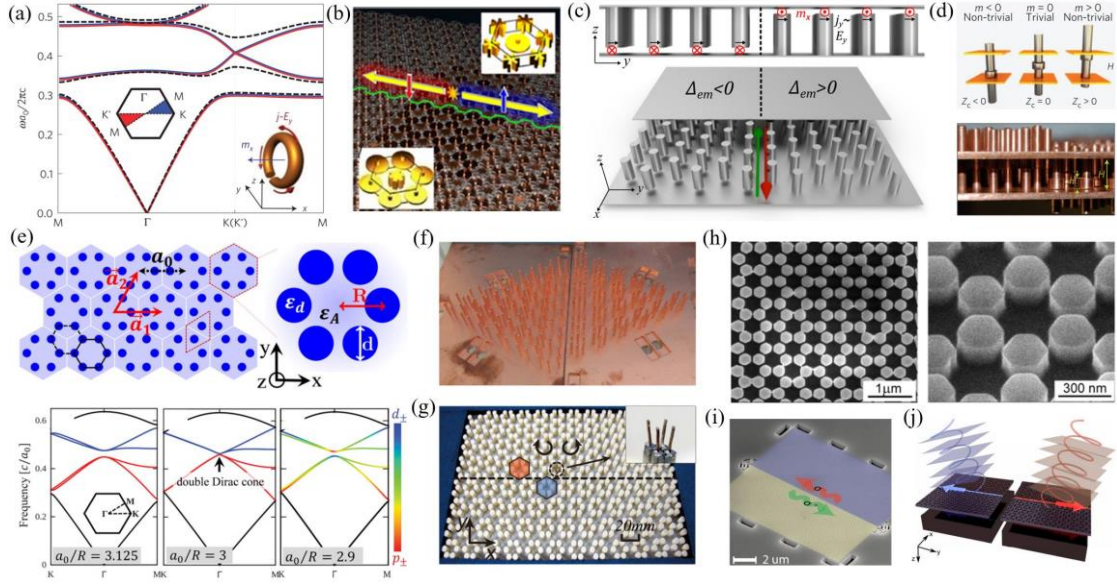

**Fig. 3.** Photonic analogues of the quantum spin Hall effect. (a) Band structure of a metacrystal with a hexagonal lattice [37]. (b) Experimental realization of QSHE by metacrystal waveguides [38]. (c),(d) Theoretical proposal [39] and experimental realization [40] of QSHE by bianisotropic metawaveguides. (e) Schematic of a triangular all-dielectric photonic crystal as a photonic analogue of QSHE [42]. (f)-(j) The experimental realization of the QSHE by crystalline metamaterials at the microwave range (f) [43] and (g) [44], visible spectral range (h) [45], near-infrared region(i) [46], and terahertz range (j) [47].

When it comes to the classical counterpart of QSHE, the time reversal symmetry for both photons and phonons are bosonic with $T_b^2 = 1$, where Kramers degeneracy is not guaranteed. To emulate the QSHE for photonic system, an essential step is to construct Kramers degeneracy by pseudospin states with pseudo- time reversal symmetry satisfies $T_s^2 = -1$. The first proposal for photonic QSHE was developed by Khanikaev et al. [37], where Kramers degeneracy is fabricated by designing metamaterials with equal effective electric permittivity tensors $\varepsilon$ and magnetic permeability tensors $\mu$, i.e., $\varepsilon = \mu$. To do so, an electromagnetic duality symmetry guarantees that the Maxwell equations are invariant under the transformation of $\varepsilon \rightarrow \mu$, $\mu \rightarrow \varepsilon$, $E \rightarrow B$ and $B \rightarrow E$, where $E$ and $B$ are electric and magnetic fields, respectively. As shown in Fig. 3(a), red and blue lines indicate the double degenerate band structure, which is achieved by two linear combinations of electric field ($E$) and magnetic field ($B$), and each of them is a pseudo- time reversal counterpart of the other. At the time-reversal-invariant points, i.e., K and K′ points in Fig. 3(a), two degenerate bands with opposite pseudospins forming a four-fold degenerate Dirac point. To open a bandgap at the Dirac point, a bi-anisotropic tensor $\chi$ (magneto-electric coupling parameter) is introduced as the physical manifestation of SOC in photonic systems. As a consequence, the four-fold degenerate Dirac point is unfolded by $\chi$, and each band is reduced to double degenerate. A non-zero $\chi$ can open a topological bandgap, as a photonic analogue of QSHE, topological interface states with spin-momentum locking can be obtained in the interface of two non-trivial domains consisting of metamaterials with opposite bi-anisotropic tensor $\chi$. However, experimental realization of the photonic analogue of QSHE is difficult, as the effective permittivity and permeability-matched ($\varepsilon = \mu$) metamaterial is hard to apply over a broad frequency range due to its high dispersive property. To overcome this drawback, a new experimental method was proposed by Chen et al. by embedding the metamaterials into a waveguide, where the permittivity and permeability matching condition can be satisfied in a broad frequency range [38]. As the experimental samples shown in Fig. 3(b), a photonic QSH system is realized by metacrystal waveguides. When combining the photonic topological insulator with a photonic ordinary insulator together, topological edge states with different propagation directions (indicated by yellow arrows), comprising spin-up (blue arrow) and spin-down (red arrow) states, respectively.

Another method to realize the photonic analogue of QSHE was theoretically proposed by Ma et al. [Fig.3(c)] [39] and experimentally realized by Cheng et al. [Fig.3(d)] [40]. The photonic version SOC is realized through the bianisotropy of metawaveguide. Specifically, when all cylinders attach to both confining metal plates, a four-fold Dirac point is constructed by two two-fold degenerate bands comprising TE and TM modes that emulate two pseudospins. When breaking the $\sigma_z$ mirror symmetry, effective bianisotropy is introduced to act as the SOC, which splits the four-fold Dirac point and opens a complete topological bandgap. With this platform, topological edge states featured by spin-momentum locking property, propagating without reflections along sharp bends of the interface, can be realized by constructing an interface with two meta-waveguides possessing opposite bi-anisotropy. Inspired by these proposals, the realization of photonic QSHE by bi-anisotropic metamaterials has also been extended to all-dielectric topological metasurfaces [41].

On the other hand, another paradigm to implement photonic analogues of QSHE via crystalline symmetries was proposed by Wu and Hu [42]. Dirac cones merge at the Brillouin zone corners K and K' points when solving the band structure of a rhombic primitive unit cell. When considering an enlarged unit cell consists of six dielectric cylinders [Fig. 3(e)], the Dirac cones fold back to the Γ point, which induces a double Dirac cone at the center of the first Brillouin zone comprising two dipole modes and two quadrupole modes. To emulate the QSHE, pseudospins are constructed by the dipole modes and quadrupole modes, while SOC is realized by shrinking or expanding the distance of cylinders, which induces a topological phase transition from photonic ordinary insulator (POI) to photonic topological insulators (PTI). Topological edge states with spin-momentum locking property can be realized at the interface between POI and PTI.

The realizations of photonic analogues of QSHE by crystalline symmetries require simple and practicable configurations in experiments, which is in favor of the extension to other frequency regimes and platforms [43]-[47], such as the experimental realization of QSHE by subwavelength resonant metal antenna [Fig. 3(f)] and all-dielectric photonic crystals [Fig. 3(g)] at microwave regime. Later, the topological edge states for QSH system were experimentally realized by a honeycomb lattice composed of nanoscale silicon Mie resonators at the visible spectral range [see Fig. 3(h)]. Robust transmission of the topological edge states in the photonic analogue of QSHE was experimentally verified with a single quantum emitter and a two-dimensional photonic crystal slab in the near-infrared region, where the strong light-matter coupling in the quantum domain was explored [Fig. 3(i)]. Subsequently, the spin-momentum locking and chiral routing along with sharp corners of topological edge states was experimentally demonstrated in a silicon photonic platform at telecom wavelengths by far-field radiation [Fig. 3(j)].

## 2.3 QVH topological metasurface
Inspired by the research advances in 2D materials, especially $MoS_2$[48] and graphene[49][50], the topological valley-Hall insulating phase has been introduced into photonics[51]-[84]. The valley refers to the local extremum of the conduction band or the valence band, usually appearing at the corners of the Brillouin zone. The excited states at each valley carry opposite angular momenta, providing valley-locked orbital magnetic moments, which are also known as valley pseudospins. Over the past few years, much attention has been given to metasurfaces with the topological valley-Hall insulating phase (topological valley-Hall metasurfaces, for short) [58][64][65][67]-[69][71]-[74][76][78][79][84][85], which exhibit many intriguing phenomena attached to valley-contrasting physics, such as the photonic valley-Hall effect, valley-locked bulk transport, and valley kink states. Exploiting the topological valley-Hall insulating phase has enabled many promising applications in robust waveguides[69][85], on-chip communications[86], and antennas[54]. In this section, we will introduce the fundamental physics and potential applications of the topological valley-Hall metasurfaces.

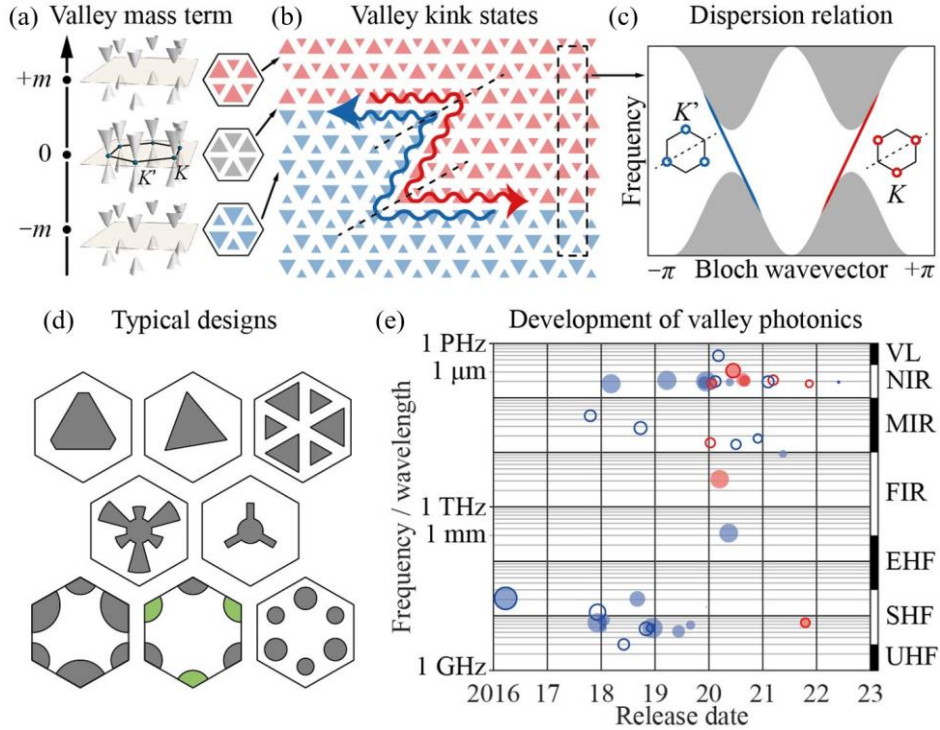

**Fig. 4.** Valley-Hall PTI and its metasurface realizations. (a) Illustration of Dirac cones with different mass terms located at the corners of the Brillouin zone. (b) Schematic diagram of domain wall consisting of two valley-Hall PTIs. (c) Dispersions of the supercell shown in (b) (dashed rectangle). (d) Typical unit cells of the topological valley-Hall metasurfaces; (e) Statistics of existing literature on topological valley photonics by June 2022. Each bubble denotes a theoretical/experimental (hollow/solid) literature on active/passive (red/blue) devices. The size denotes the number of citations.

Different from photonic Chern insulators with broken time-reversal symmetry[87]-[89] and spin-Hall photonic topological insulators (PTIs) protected by time-reversal symmetry[42][44][90], the valley-Hall PTIs require the breaking of the inversion symmetry ($P$)[53]. Considering a graphene-like photonic crystal featuring two Dirac cones, when perturbatively breaking $P$, the pair of Dirac cones will be lifted, and a bandgap opens at K/K' valley. The band structure around K/K' valley can be described by a massive Dirac equation, with a mass term $m$, whose magnitude measures the bandwidth and sign is determined by the nonzero local Berry flux at K/K' valley (see Fig. 4(a)). In addition, the topological property comes from the nonzero local Berry curvatures, defined by

$$\Omega_n(k) = \frac{\partial A_y}{\partial k_x} - \frac{\partial A_x}{\partial k_y}$$

where $\mathbf{A}_n = -i\langle u_n|\partial_k|u_n\rangle$ is the Berry connection with $u_n$ being the periodic part of the Bloch wave function of the nth band. Different from the Chern number obtained by integrating the Berry curvature over the entire Brillouin zone, the topological invariant of valley-Hall PTIs, known as the valley-Chern number, is calculated by integrating the local Berry curvature around the valleys. When the P-breaking perturbation is small, a valley-Chern number takes approximately the value of ±1/2[91].

A remarkable property of the valley-Hall PTI is that at the interface consisting of two valley-Hall PTIs with opposite valley-Chern numbers, dubbed the *domain wall* (see Fig. 4(b))[49], a pair of gapless boundary modes (also known as valley kink states) appear. Due to the preserved *T* symmetry, the valley kink states bounded to different valleys have opposite group/phase velocities, as shown in Fig. 4(c). Owing to their valley-locked chirality[53], the valley kink states can only scatter to themselves and transfer the energy among the same K/K' valleys in the Brillouin zone, even if the domain wall is bent by 120°. Hence, the valley kink states can go nowhere but pass through the sharp corners, implying its robustness to certain types of disorders. Up to now, different kinds of unit cells with broken *P* have been designed to realize valley-Hall PTIs. This can usually be achieved by lowering the lattice symmetry from $C_{3v}$ or $C_6$ to $C_3$, in a triangular or honeycomb lattice, as shown in Fig. 4(d).

The study of topological valley-Hall metasurfaces is not only a scientific curiosity but has also led to many promising applications. Having introduced the fundamental physics of the topological valley insulating phase, herein, we will focus on the development of topological valley-Hall metasurfaces from the perspective of practical applications. In general, photonic devices can be classified as passive and active based on whether additional pump sources or active control is required. In Fig. 4(e), we present the statistics of existing literature on topological valley photonics by June 2022. As one can see, there are two tendencies in the recent five years: i) from microwave to optics; ii) from passive to active. In this subsection, before proceeding to the active devices based on valley-Hall PTIs that involve gain or dynamic control, we will firstly review the passive devices.

High-speed on-chip communications are of the essence in modern information and communication technologies, including optical integrated circuitry and on-chip interconnect, which require high-efficiency, low-loss, integrated, and robust solutions to waveguiding. Comparing to the conventional photonic crystal waveguides or other nontopological optical waveguides, the topological valley kink states hold great potential for on-chip communications, owing to their intrinsic properties, including topological robustness against defects (determined by the band topology), single-mode propagation (guaranteed by the valley-Chern number difference at domain walls), and linear dispersion (governed by the low-energy Dirac equation). Applying the valley-kink states to on-chip communications has been experimentally demonstrated at terahertz (THz) frequencies recently[86]. As shown in Fig. 5(a), the designed THz on-chip valley-Hall PTI has a graphene-like lattice, where an array of triangular holes are patterned on a high-resistivity suspended silicon metasurface with low absorptive losses. As expected, the valley kink states supported by the silicon metasurface are topologically robust and can pass through multiple sharp bends (five 120° bends and five 60° bends) with almost unity transmission. Moreover, an error-free transmission with a data transfer rate up to 11 GBits/s around 0.33 THz was also demonstrated experimentally, which further enables real-time transmission of uncompressed 4K high-definition video. The robust on-chip propagation of valley kink states has also been demonstrated at the optical frequencies

[67][69], and the corresponding on-chip optical communications require further experimental explorations.

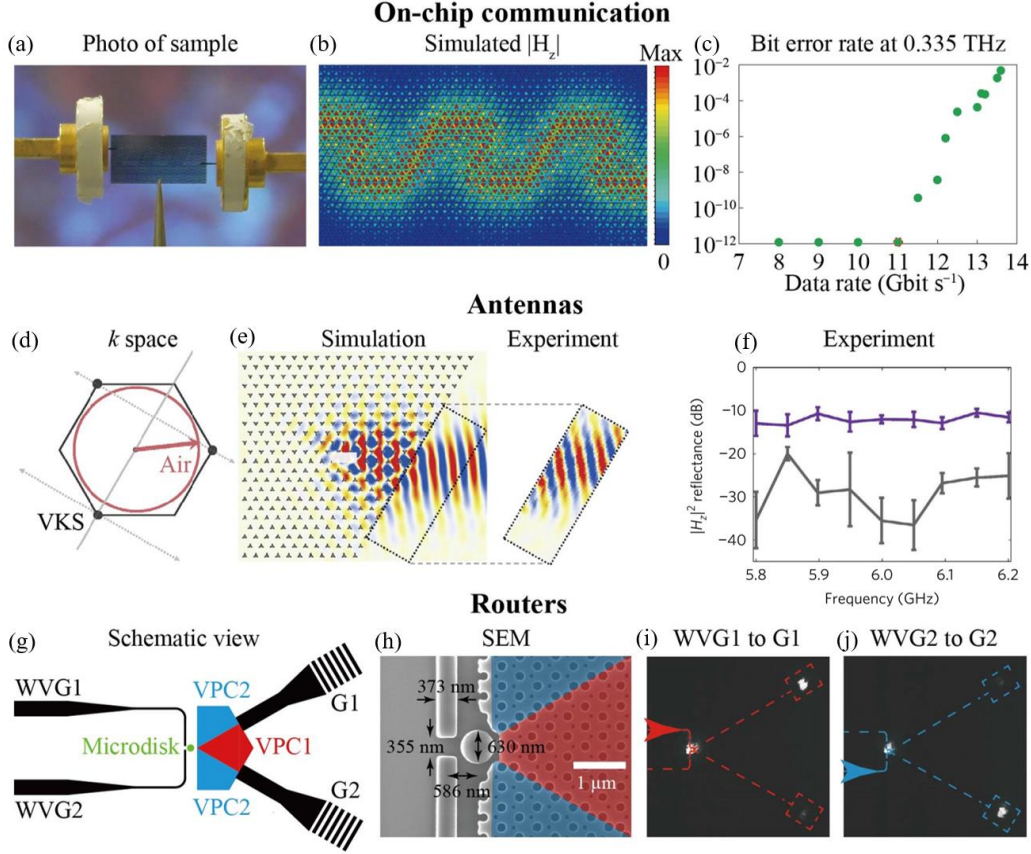

**Fig. 5.** Passive photonic devices based on topological valley-Hall metasurfaces. (a) Silicon topological valley-Hall metasurface for on-chip THz communication [86]. (b) Simulated field distribution for structure in (a) [86]. (c) Measured bit error rate as a function of data rate at 0.335 THz [86]. (d) The momentum-space analysis on the outcoupling of (e) simulated (left) and measured (right) field patterns for the outcoupling of TE-mode valley kink state to vacuum space [54]. (f) Measured reflectance for zigzag (grey) and armchair (purple) terminations [54]. (g) Schematic diagram of photonic routing based on the valley kink states [67]. (h) Scanning-electron-microscope (SEM) view of the experimental sample [67]. (i), (j) Measurement of photonic routing profiles at λ = 1400 nm for light injected from the WVG1/WVG2 port [67].

Apart from the on-chip communication, the valley kink states can also offer the possibility for high-efficiency coupling to the surrounding environment with a certain type of terminations, which can be used to devise reflectionless directional antennas[51][54][72]. Figs. 5(d)-(f) show the experimental demonstration of topological refraction of valley kink states to ambient space[54]. In experiments, a phased array of dipole antennas was placed at the centre of the domain wall (white rectangle in Fig. 5(e)) to selectively excite rightward valley kink states. The excited kink states then outcouple to the ambient space through a valley-preserving zigzag termination with negligible reflection that usually require joudicious design of impedances in the conventional out coupler (see Fig. 5(e) and the purple curve in Fig. 5(f)). The refraction direction can be understood according to the momentum phase-matching conditions at the terminal interface (see Fig. 5(d)). In contrast, as the valley conservation is broken at the armchair termination, the photonic energy is inevitably coupled to the backward channel, resulting in strong reflection (the grey curve in Fig. 5(f)). The high-efficient outcoupling from the valley kink states to ambient space may find potential applications such as directional antennas, lasers, and wireless communications.

Owing to the property of valley-locked chirality, the valley kink states can be used to devise all-optical routers operational at telecommunication wavelength, which was realized recently[67]. The experiments were performed in a silicon-on-insulator (SOI) platform, as shown in Figs. 5(g), (h). The photonic router consists of two valley domain walls connected with two nonuniform grating couplers and a microdisk connected with two 373-nm-width strip silicon waveguides (labelled WVG1 and WVG2 in Fig. 5(g)). The close-to-diffraction-limited microdisk serving as a phase vortex generator transforms the waveguide mode from WVG1/WVG2 input waveguides to a clockwise/counter-clockwise vortex field. Owing to the valley-chirality locking, the vortices with opposite chiralities are converted to different valley kink states with opposite valleys, and then couple to free space via the grating couplers. Besides, the valley-chirality locking also enables the topological channel intersection[72], where the valley transport path depends solely on the geometries of the intersection.

## 2.4 Floquet topological metasurface

Floquet topological insulators (FTIs)[92][93], a class of time-varied systems hosting robust edge states against disorders, have attracted intense research interests across multi disciplinaries from condensed matter physics to photonics. The dynamics of FTIs are governed by evolution equations $\psi(t') = U(t',t)\psi(t)$, where $U$ is the evolution operator and $\psi$ is the wave function. Basically, revealing the evolutions stands on top of the understandings on periodic modulations $U(T,0) \equiv e^{-iH_FT}$, where $T$ is the temporal period of the modulation, and $H_F$ represents the effective Floquet Hamiltonian[94]. Unlike band diagrams of static Hamiltonian systems bounded by ground states, $H_F$ exhibits periodic quasi-energy band diagrams $\varepsilon_k$-$k$, with corresponding eigen states $\varphi_k(t)$ satisfying $\varphi_k(t) = \varphi_k(t+T)$. Generally, $H_F$ can be generally written as $H_F = \vec{n}_k \cdot \vec{a}_k + \varepsilon_k I$, where $\vec{a}_k$ and the unit matrix $I$ form an orthogonal basis for $N \times N$ matrices in a Hilbert space. The bulk topologies of FTI are characterized with Chern number, $C_F = \frac{1}{4\pi}\iint_{BZ} d^2k (\partial_{k_x}\hat{n}_k \times \partial_{k_y}\hat{n}_k) \cdot \hat{n}_k$, where BZ represents the Brillouin zone, and $\hat{n}_k = \vec{n}_k/|\vec{n}_k|$. Intuitively, fruitful Floquet topological phases can be anticipated from distinct modulations, which is challenging to realize in condensed matter systems[95][96]. Due to the easy implementations and flexibilities in wave manipulations, photonic metasurfaces promise in demonstrating FTI phases. On the other hand, exotic FTI phases offer novel approaches to manipulate light, thus promising unusual photonic applications.

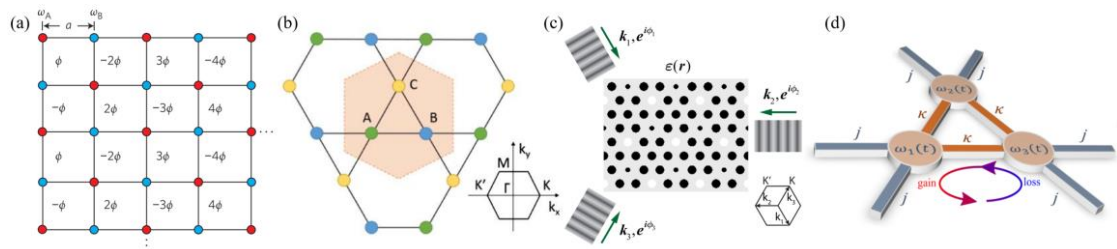

**Fig. 6.** (a) Dynamically modulated photonic resonator lattice exhibiting an effective magnetic field for photons [97]. (b) Kagomé lattice with three sites in the primitive cell, and the corresponding Brillouin zone [98]. (c) The FTI consists of a static PhC and the permittivity modulations by three Bloch waves [99]. (d) Non-Hermitian Floquet kagomé lattice [101].

Theoretically, fruitful Floquet topological phases have been predicted on various time-variant photonic systems [97]-[134], including Floquet Chern insulators (FCTIs) [97]-[101], and anomalous Floquet topological insulators (AFTIs) [101]. Regarding the FCTIs, whose bulk bands are characterized by non-zero Chern numbers, two types of models (i.e. discrete and continuum) have been proposed respectively. In 2012, A square lattice of photonic resonators with synthesized Landau gauge has been proposed to break time-reversal symmetry by Fang *et al* [97]. In that lattice, the Landau gauge is synthesized by applying harmonical modulation on coupling coefficients, which give rise to modulation phases. The specific distribution of modulation phases is shown in Fig. 6(a), where all the horizontal couplings are in-phase, but the vertical couplings are different. Such phase distribution ensures that light passing through a plaquette will accumulate a constant phase $\varphi$, which is equivalent to the gauge field. The effective gauge field breaks the time-reversal symmetry of the system, leading to non-zero Chern number $C_F = 1$, manifesting as a single chiral edge state. Such aperiodic coupling scheme requires precise regulations on modulation phases, therefore may pose challenges in experimental realizations. Avoiding the aperiodic restriction, Minkov *et al*. proposed a periodic FCTI based on Haldane model[98][102]. In a Kagomé lattice of resonators, the resonant frequencies are time-periodically modulated as shown in Fig. 6(b). According to perturbation theory, the effective Floquet Hamiltonian can be expanded in orders of $1/\Omega$ as $H_F = H_{0\Omega} + H_{1\Omega} + \mathcal{O}(1/\Omega^2)$, where $\Omega$ is the frequency of modulation. In this system, the first-order term $H_{1\Omega}$ is purely imaginary, which corresponds to the second-neighbor complex coupling in Haldane model. Compared with Ref. [97], the system simply consists of an array of identical and single-mode resonators without intermediate resonators. In addition, three mechanisms have been proposed for such dynamic modulation on frequencies: electro-optic modulation, optomechanical modulation and optically induced Kerr nonlinearity. Unlike discrete models based on resonators and couplings, continuum models are proposed to realize FTIs on top of concrete dielectric media[99][100]. Fang *et al*. proposed a linear modulation scheme to synthesize effective magnetic fields in photonic crystals (PhCs)[99]. The dielectric permittivity is modulated with multiple Bloch elastic waves, which generate distributions of modulation phases. In addition, they reported that in a triangular lattice with honeycomb sublattice (Fig. 6(c)), when the Bloch-wave has the same periodicity as the static PhC, the net effective magnetic flux through a unit cell vanishes, but the Floquet bands attain non-zero Chern numbers. Besides the linear modulatoin, a nonlinear modulation scheme has also been proposed. He *et al*. reported a FCTI in nonlinear PhCs by engineering the external drive to break time-reversal symmetry (*T*)[100]. The Floquet band gaps can be closed and re-opened by engineering the driving field (polarization and frequency). Besides, they proposed a FCTI by breaking *T* using elliptically polarized driving fields in a hexagonal lattice of silicon and *z*-cut LiNbO$_3$.

In AFTIs, robust chiral edge states appear even though the Chern numbers of all the bulk Floquet bands are zero, and the topological invariants characterizing AFTIs are winding numbers. Li *et al.* studied the topological phases and states in non-Hermitian FTIs. The structure is a Kagomé lattice with nearest-neighbor hopping[101] as shown in Fig. 6(d), and the non-Hermitian periodic modulation is introduced by adding complex time-dependent perturbations to the on-site frequencies. The numerical results reveal that the system presents conventional FCTI phase with weak perturbations, and AFTI phase with strong perturbations. Topologically protected edge states exist in both phases.

Due to the temporal modulation itself in either optical or mechanical systems is a rather challenging task, therefore, many proposals remain theoretical. By contrast, effective modulation[103]-[108]creates the possibility of a practical implementation of FTIs. Currently, effective modulations are all based on coupled ring resonator systems. The method of effective modulation was theoretically proposed by Liang *et al*. [103][104], and then was demonstrated in microwave[105][106] and optical[107][108] regions subsequently. In 2013, Liang *et al*.[103] considered that clockwise and counter-clockwise propagation modes in ring resonators through a plaquette acquire effective gauge potentials with opposite signs. the system enters the AFTI phase and exhibits one-way edge states, with non-zero $Z_2$ topological invariant. Subsequently, Pasek *et al*. pointed out that such resonator-based photonic FTI can be modeled as networks[104], as shown in Fig. 7(a). The Bloch modes of periodic network models can be mapped onto the Bloch-Floquet states of driven lattices. Besides, network models based on the honeycomb lattice have richer phase diagrams, including FCTIs and AFTIs phase under different coupling strengths.

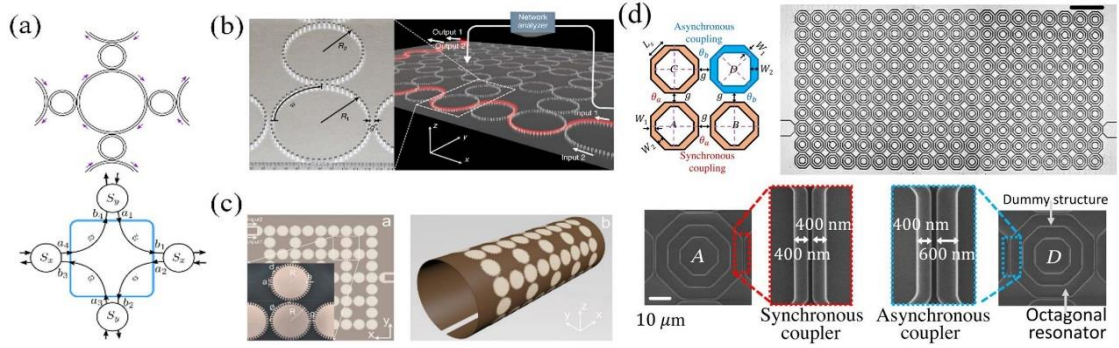

**Fig. 7.** (a) Schematic of a unit cell in a 2-D lattice of photonic ring resonators (upper) and the equivalent periodic network (lower) [104]. (b) Photo of metallic rods on a flat metallic surface (left), and the schematic of a 5×5 lattice in experiment (right) [105]. (c) photo of the CSP ring resonators printed on a flexible paper-like dielectric film (left) and the schematic of the folded flexible photonic TI (right) [106]. (d) Schematic of a unit cell of a Floquet lattice of identical, evanescently coupled octagon resonators, with octagon D rotated by 45° with respect to the other three resonators [107].

The AFTI phase based on the network model was firstly demonstrated by Gao *et al.* in 2016[105]. They experimentally implemented the square lattice in Ref. [103] by using designer surface plasmon (SP) structures operating in the microwave regions. Fig. 7(b) shows that the sample consists of closely spaced sub-wavelength metallic rods, placed on a flat metallic surface. Large rings (lattice rings), are set in a square lattice, and each pair of adjacent lattice rings is connected by a smaller ring (coupling ring). Based on this design, they discussed the robustness of AFTI against a variety of defect classes. Such design further goes to flexible scenarios. Exploiting the same mechanism, Gao *et al.* reported the experimental realization of a paper-like flexible AFTI by using conformal SP (CSP) and standard printed circuit board (PCB) technology[106], as shown in Fig. 7(c). Due to the flexibility of the fabricated FTI, spatial topologies can be constructed via folding and connecting, chiral topological edge states are demonstrated in both scenarios.

The first experimental realization of an FTI on a nanophotonics platform was reported[107]in 2020. As shown in Fig. 7(d), their sample is a square lattice of strongly coupled octagonal resonators in the silicon-on-insulator (SOI) material system, and the resonators A and D have different coupling coefficients with their respective adjacent resonators. As light circulates around each microring, it interacts periodically with its neighbors with different coupling coefficients, the lattice can thus being equivalent to a periodically driven Floquet system. Besides, by exploiting the frequency dispersion of the evanescent couplers to adjust the coupling strength, they experimentally demonstrated the topological phase transition between FCTI and AFTI. The realization of on-chip FTI has also spawned applications.

To date, the FTIs based on dynamic modulations have not been experimentally demonstrated, due to the extremely weak electro-optical and nonlinearity of materials, especially at high frequencies. On the other hand, the effective modulations have been experimentally demonstrated, however they are so far limited in coupled ring resonator scheme. Straightforwardly, more efforts would be needed on advanced experimental technologies for dynamically-modulated FTI and effective-modulation schemes beyond coupled ring-resonators array. Besides, more intriguing physics could be anticipated by connecting the FTI with other mechanisms, i.e., non-Hermicity[109][110] and quantum optics[111][112].

## 2.5 High-order topological metasurfaces

Different from the conventional topological states where the topological boundary states appear at 1 dimension less than the bulk, second corner states are higher-order topological phenomena where the topological boundary states appear in dimensions at least second order lower than the bulk [113][114]. This new topological phase has been explored in different photonic structures and has found diverse applications, which we will discuss in this section.

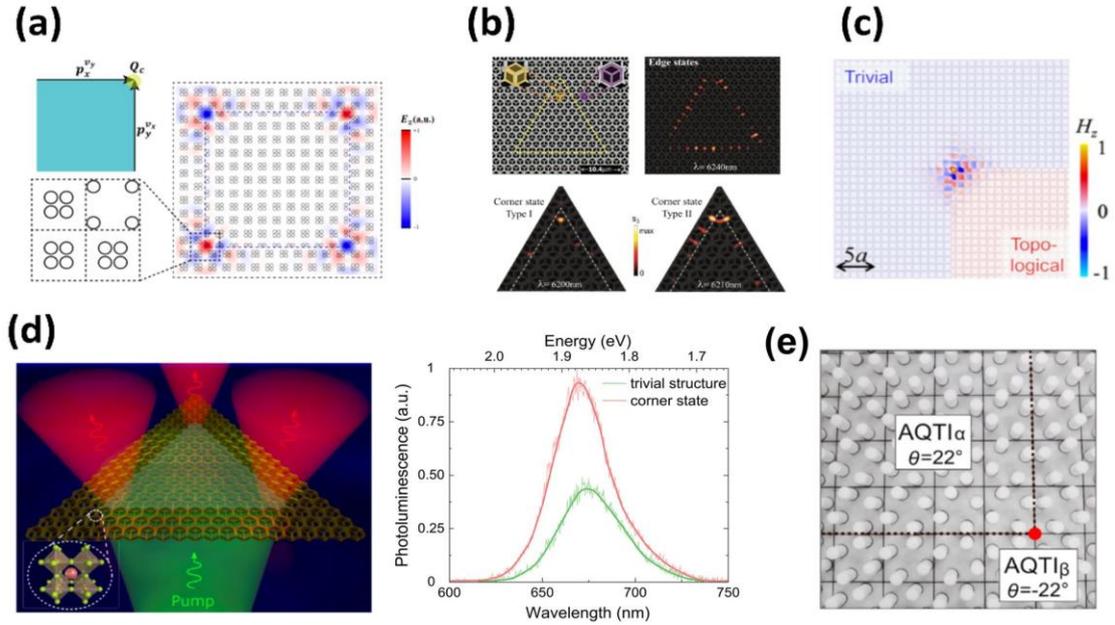

**Fig. 8.** Second-order photonic corner states. (a) Second-order photonic corner states in a photonic crystal with dielectric rods [115]. (b) Second order photonic corner states in a Kagome metasurface [120]. (c) Photonic crystal nanocavity based on a topological corner state [121]. (d) Enhanced photoluminescence mediated by a topological metasurface [123]. (e) Quadrupole topological phase in a twisted photonic crystal [130].

The most popular models on topological corner states in photonic systems are based on the 2D SSH model, in which each unit cell contains some cylinders and by expanding or shrinking the cylinders' positions away or towards the center of the unit cell, intra-cell and inter-cell hopping could be tuned, leading to a topologically trivial to nontrivial transition. For the square lattice 2D SSH model [115], as shown in Fig. 8(a), it considered a 2D photonic crystal with four identical dielectric rods in each unit cell and by adjusting the distances between the nearby rods in the x and y directions, the emergence of edge and corner states can be controlled straightforwardly. The 2D square SSH model was experimentally realized in photonic crystal slabs with periodic dielectric rods on a perfect electric conductor [116] and in photonic crystals consisting of alumina cylinders sandwiched between two metallic plates [117]. Moreover, based on the coupled dipole approximation, the polarization-dependent topological phase of an array of metallic nanoparticles in 2D SSH lattice has been studied [118], showing that the polarization can be served as a new degree of freedom to control the topological features and develop robust multifunctional photonic devices.

Corner states have also been studied in Kagome lattice whose unit cell contains three cylinders. Especially, corner states in Kagome lattices have been experimentally observed in array of dielectric cylinders arranged to form a Kagome lattice between two parallel aluminium plates [119], and in metasurface fabricated on a silicon-on-insulator chip consisting of trimers of diamond-shaped holes [120], as demonstrated in Fig. 8(b). Interestingly, in addition to the corner states due to nearest-neighbour interactions, a new class of topological corner states induced by long-range interactions with a purely electromagnetic nature, which has no analogy in condensed-matter systems, exists in the Kagome structure.

Corner states can serve as high-Q cavity modes, thus providing potential applications in enhancing light-matter interaction. As shown in Fig. 8(c), a corner state tightly localized in space with a high Q factor over 2000 was experimentally observed in [121], verifying its promise as a nanocavity. The corner states could be pumped using in-plane excitation conditions as experimentally demonstrated in [122]and could be used for enhancement of photoluminescence signal [123] (see as depicted in Fig. 8(d)), and rainbow trapping [124]. For example, multiband corner states have been designed via the topological optimization method [125], in which the system supports four highly localized corner states within four sizeable band gaps that are robust to bulk impurities. The recently proposed dual-polarization topological corner states for both TE and TM modes [126] and a new principle for creating corner states within odd-order band gaps in $C_{4v}$-symmetric lattices beyond the 2D SSH paradigm [127] open new possibilities for both fundamental science and promising applications.

While corner states could emerge due to nontrivial bulk dipole moment as discussed above, they can also appear due to nontrivial bulk quadrupole moment without dipole moment. This kind of insulator was called quadrupole insulator in the literature and is challenging to implement using photonic systems due to the existence of negative couplings in the original pi-flux model. Up to now, there are only a few works investigating quadrupole topological states in photonic systems [29][128]-[130]. In [128], the authors experientially implemented the negative coupling in a 2D lattice of nanophotonic silicon ring resonators and demonstrated that the quantization of the bulk quadrupole moment manifests as topologically robust 0D corner states. Negative coupling could also be introduced in a lattice of plasmon-polaritonic nanocavities exploiting the geometry-dependent sign reversal of the couplings between the daisylike nanocavities [129] or in gyromagnetic photonic crystal through a double-band-inversion process [29]. More interestingly, quadrupole topological phases could be realized in all-dielectric photonic crystals without the pi-flux-threading mechanism as experimentally demonstrated in photonic crystals composed of dielectric cylinders via twisting the unit-cell [130], as shown in Fig. 8(e).

The dispersive topological edge states for QSHE in photonic topological insulator can be used to realize topologically protected mid-gap defect modes (or corner modes) by opening a band gap for the edge state. Topological corner modes with small mode volume are robust against structure deformations, which was first observed by Noh et al. in a femtosecond-laser-written waveguide array [131]. The photonic analogue of high-order QSHE is experimentally realized in all-dielectric photonic crystals by observing directional localization of speudospin-polarized corner states [132].

Based on Floquet HOTIs, topological states with dimensions two or more lower than that of the bulk can also be studied in Floquet topological metasurfaces. Connecting with the concept of synthetic dimensions, Dutt *et al.* proposed photonic HOTI by using dynamic modulation[133]. In a 1-D chain of ring resonators, the two rings in a cell are anti-symmetrically modulated at the frequency spacing between the ring modes, a quadrupole HOTI in the synthetic frequency dimension is constructed, which hosts topologically nontrivial corner modes. Based on the synthetic dimension, a proposal on a mode-locked topological laser has also been reported[134].

## 2.6 Other topological metasurface
### 2.6.1 BIC

Bound states in the continuum can completely confine light by eliminating radiation loss even though their frequencies and momentum are embedded in the continuum spectrum[135]. BICs were first mathematically proposed in 1929 by von Neumann and Wigner in quantum mechanics with artificial quantum potential[136]. BICs were subsequently realized as destructive interference[137] and attracted broad attention in both quantum and classical waves. They can be roughly classified as symmetry-protected BICs and off-Γ BICs. Symmetry-protected BICs[138]-[140] are realized when there is a symmetry mismatch between resonances and radiation channels. Off-Γ BICs can be realized by various mechanisms. For example, Friedrich-Wintgen BICs (FW-BICs) are generated when there is destructive interference between multiple resonances[137][141][142]. Tunable BICs can be realized when there is an accidental cancelation of radiation loss at one single band[143]. Radiation channels can be reduced by environmental design to achieve BICs[144]. BICs with an infinite lifetime can improve light confinement and have shown benefits in boosting light-matter interactions.

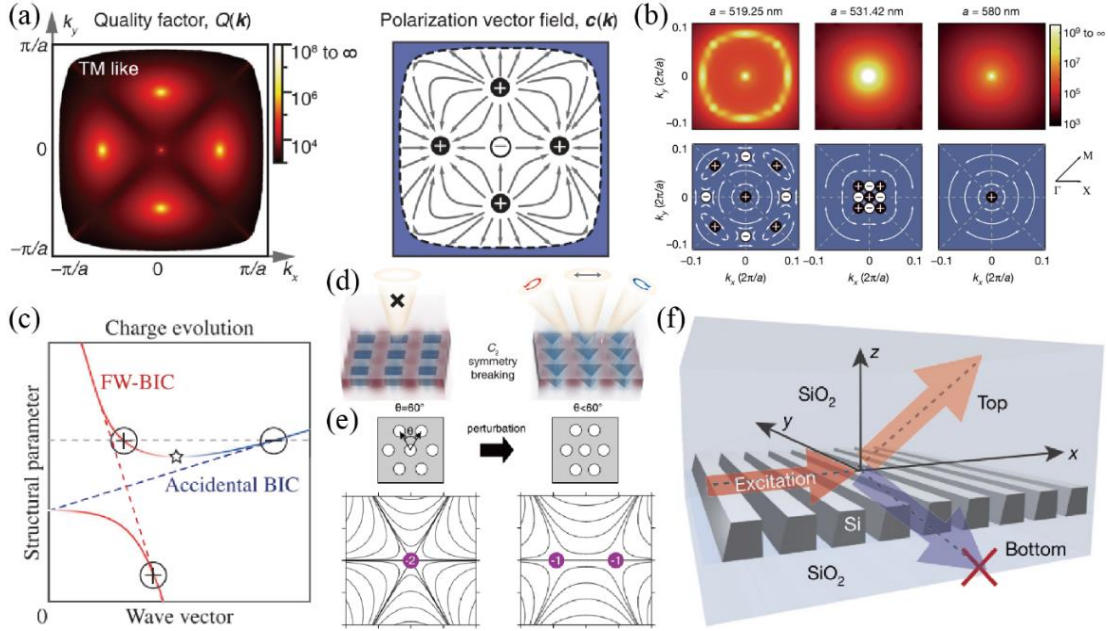

**Fig. 9.** (a) Far-field polarizations form vortexes with BICs as polarization singularities [145]. (b) Multiple BICs are tuned to gather together as a merging BIC. The Q factors of nearby resonances have been significantly enhanced at the merging BIC compared to isolated BICs [148]. (c) The accidental BIC and the FW-BIC are tuned to merge at an off-Γ point by varying structural parameters [149]. (d) Circularly polarized states are spawned from BICs under $C_2$ symmetry breaking. Adapted with permission[151]. (e) The higher-charged BIC is split into two off-Γ BICs by reducing symmetry [152]. (f) UGR are created

by eliminating radiation loss using polarization singularity at only one single side [153].

BICs have been recently revealed as topologically protected polarization singularity in the momentum space[145]. The polarizations of far-field radiation evolve and form vortexes around a BIC, as shown in Fig. 9(a). The winding number of polarization defines the topological charge of a BIC. BICs with undefined polarization have vanishing coupling with radiation channels and therefore are confined with an infinite Q factor. The topological nature has been theoretically proposed by Zhen[145] et al. in 2014. Subsequently, experimental verification has been separately performed by Doeleman et al. in a one-dimensional grating[146] and Zhang et al. in two-dimensional periodic plasmonic structures[147]. BICs carry integral topological charges and are topologically protected by topological charge conservation. The dynamics of topological charges including evolution, merging and split in the momentum space has enabled interesting applications.

When multiple BICs are tuned to the same point, a merging BIC is created, which is promising in designing a super-high Q cavity. Although BICs have infinite Q factors theoretically, there are inevitable fabrication imperfections that can couple BICs with nearby radiative states through scattering, which limits the available Q factor. To further improve the light confinement, Q factors of nearby radiative states need to be enhanced. As for an isolated BIC, the scaling rule of Q factors decaying away from the singularity has been limited by its topological charge. Recently, merging BICs have been proposed by Jin et al. to significantly improve the Q factors of nearby states[148]. They have constructed a merging BIC by gathering off-$\Gamma$ BICs together with a symmetry-protected BIC at the $\Gamma$ point, as shown in Fig. 9(b). Off-$\Gamma$ BICs are sensitive to geometric perturbation and tunable with momentum selection. Experimental demonstration has been accomplished to prove that the Q factor with robustness has been enhanced over one order of magnitude. Subsequently, Kang et al. have done a further study to realize merging BICs at off-high symmetry points by merging two different mechanisms induced off-$\Gamma$ BICs, i.e., accidental BICs and FW-BICs[149], as shown in Fig. 9(c). The two BICs are simultaneously realized at the same branch of two coupled resonances. They carry opposite topological charges and can be tuned to form a merging BIC. By further reducing symmetry, merging BICs are tunable in the momentum space. Recently, Kang et al. have proposed another scheme for constructing merging BICs by manipulating higher-charged BICs[150]. When the topological charge involved in a merging BIC has been increased, the Q factor of nearby radiative states can be further enhanced. Moreover, higher-charged BICs can be split into multiple BICs by reducing symmetry and can be constructed merging BICs with other off-$\Gamma$ BICs.

When BICs are broken by reducing symmetry, the topological charge cannot disappear and will split as smaller topological charges following topological charge conservation. Liu et al. have demonstrated that circularly polarized states (C points) can emerge from the elimination of a BIC[151], as shown in Fig. 9(d). An integral topological charge is broken by the in-plane inversion symmetry ($C_2$) breaking and split as two half-charges with C points as polarization singularity. Elliptical polarizations evolve around the C point and can cover the whole Poincaré sphere. On the contrary, C points with the same topological charge can be tuned to merge as a BIC. Higher-charged BICs can appear when there is a higher rotation symmetry. When the rotation symmetry is reduced, the higher charge is no longer allowed and split as lower charges. As demonstrated by Yoda et al., when they broke the $C_6$ rotation symmetry protecting a BIC with topological charge -2 and kept the $C_2$ symmetry, two off-$\Gamma$ BICs with topological charge -1 are generated[152], as shown in Fig. 9(e). When the $C_2$ symmetry is further broken, paired C points with the same half topological charges and opposite handedness are generated. The upward and downward radiation are associated with the up-down mirror symmetry and therefore can be eliminated simultaneously. Yin et al. have proposed and experimentally demonstrated that unidirectional guided resonances (UGR) with vanishing radiation only on a single side can be created by breaking the up-down mirror symmetry[153] as shown in Fig. 9(f). Paired C points with the same half topological charges emerge from a broken BIC. When they are tuned to merge at a single side, a polarization singularity with eliminated radiation is generated, while radiation remains on the other side. Paired C points with opposite half topological charges can be created and constructed to form UGR[154], as demonstrated by Zeng et al.

BICs with infinite Q factors are promising in boosting the performance of light propagation[155]-[160]. Zero-index materials have been designed in all-dielectric photonic crystals with Dirac-cone dispersion to alleviate ohmic losses. However, they are suffering from out-of-plane radiation loss that hinders their development. BICs have been introduced to completely eliminate radiation loss. As theoretically demonstrated by Minkov et al., Dirac-cone dispersion has symmetry-protected BICs at the degenerate[157]. When another single band with a symmetry-protected BIC is tuned to the degenerate, a zero-index material is realized, as shown in Fig. 10(a). Subsequently, Dong et al. have realized BICs in a zero-index material using resonance trapped modes with destructive interference radiation[158]. Tang et al. have experimentally implemented the low-loss zero-index material[159]. In addition, BICs have been explored in other applications to remove radiation loss during light propagation[155][156][160]. For example, diffraction-free beams realized at a BIC have been demonstrated to confine beams in out-of-plane[160], as shown in Fig. 10(b).

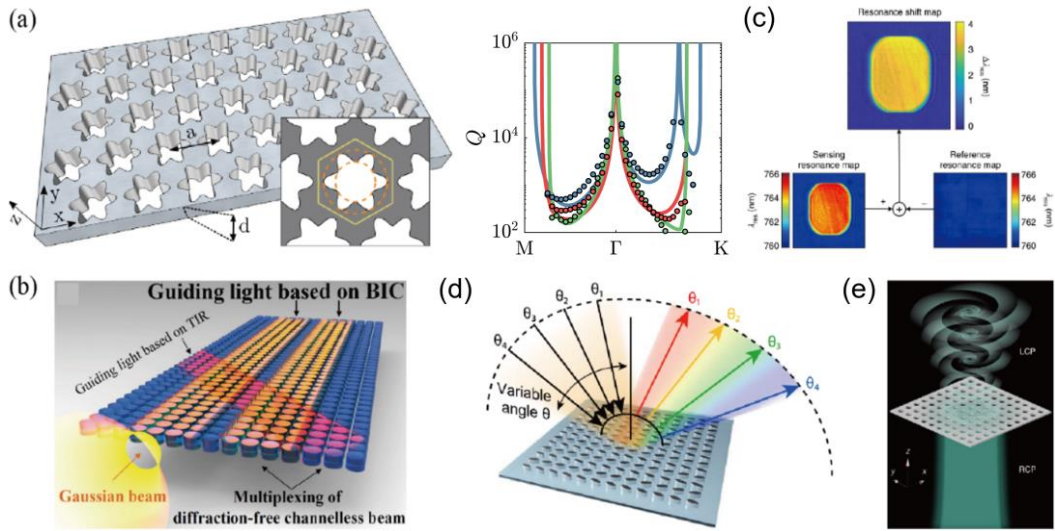

**Fig. 10.** (a) Zero-index materials with light confined by BICs in out-of-plane [157]. (b) Diffraction-free beams are guided by BICs beyond the light cone [160]. (c) Ultrasensitive hyperspectral imaging by detecting BIC-inspired resonance shifts [163]. (d) Angular-scanning sensors using BIC-inspired narrow spectra [164]. (e) Optical vortexes generated from the polarization vortex around BICs [167].

Metasurfaces supporting BICs possess ultrasharp resonances and strong light confinement, which can be extremely sensitive to refractive index changes[161]-[166]. Chemical or biological sensing enhanced by BICs has been explored recently. Yesilkoy et al. have developed ultrasensitive label-free biosensing by combining quasi-BICs in all-dielectric metasurfaces and hyperspectral imaging[163], as shown in Fig. 10(c). Resonances in metasurfaces have a perturbation with symmetry-protected BICs and are tunable with the size of paired tilted silicon nanobar. Once the local refractive index has been changed by analytes, the resonance shift compared with the reference appears and composes hyperspectral imaging. Leitis et al. have developed angular-scanning sensors to detect molecular absorption fingerprints using BICs-inspired narrow spectra[164], as shown in Fig. 10(d). When the resonances at one incidence angle match with the vibrational modes of analytes, a strong modulation appears in the reflection. Molecular absorption fingerprints can be retrieved from the deviation in angle-resolved reflection. Chen et al. have integrated chiral sensing and refractive index onto an individual metasurface and utilized quasi-BICs to improve the sensitivity[165].

The topological property of polarizations winding around BICs provides a feasible approach for generating optical vortexes. As demonstrated by Wang et al., when circularly polarized light is incident around a BIC, the transmitted light will gain a Pancharatnam-Berry phase[167]. The Pancharatnam-Berry phase changes with the winding polarizations in the momentum space. As a result, under the shining of a slightly convergent beam at the iso-frequency contour, the transmitted beam carries a spiral phase front, as shown in Fig. 10(e). The topological charge of the vortex beam is determined by the topological charge of the BIC. This novel approach has no requirement for the intricate design of inhomogeneous metasurfaces and therefore there is no need for accurate alignment to the incident beam center.

**2.6.2 Skyrmions**

Skyrmions are topologically stable quasiparticle excitations initially proposed as a nucleon model in 1961 [168]-[170]. Skyrmions have been predicted and investigated in liquid crystals [171], Bose-Einstein condensates [172], and are most commonly known in thin-film magnetic materials [173]-[178] due to their promising potential in spintronics. Recently, skyrmions and related concepts have also drawn much attention in electromagnetic waves. Here same as before, we restrict our review to 2D systems and briefly introduce the recent advances of skyrmion-related concepts in light.

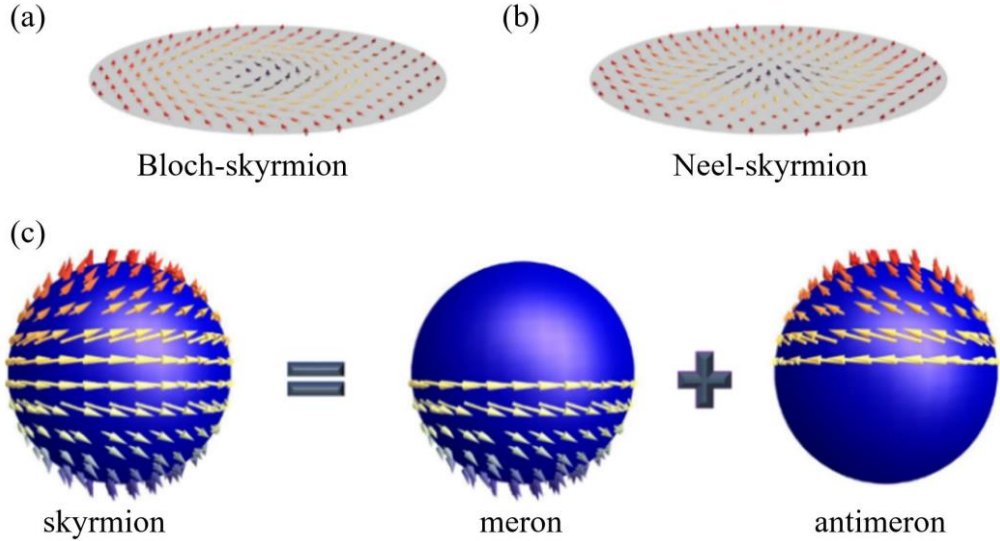

**Fig. 11.** (a) a Bloch type skyrmion and (b) a Néel type skyrmion with $p = 1$ and $m = 1$. (c) An illustration shows the relation between a skyrmion and a meron (antimeron).

Skyrmion-related objects define topological nontrivial three-component vector fields (denoted by $\boldsymbol{n}$) over a 2D coordinate space ($x, y$). The skyrmion number can characterize the topology of this nontrivial vector field distribution [179][180],

$$N_{sk} = \frac{1}{4\pi} \iint dxdy\, \boldsymbol{n} \cdot \left(\frac{\partial \boldsymbol{n}}{\partial x} \times \frac{\partial \boldsymbol{n}}{\partial y}\right). \tag{1}$$

In typical situations, the magnitude of $\boldsymbol{n}$ is irrelevant, and thus $\boldsymbol{n}$ forms a two-sphere $S^2$. Meanwhile, skyrmions as quasiparticle excitations are assumed to be embedded in a uniform background and thus $\boldsymbol{n}$ at the outskirt of the skyrmions are the same. One can merge the edge of the skyrmion into a single point, and then the 2D coordinate space also forms a two-sphere $S^2$. The skyrmion number defines a map $S^2 \rightarrow S^2$ characterized by the second homotopy group of the sphere $\pi_2(S^2) = \mathbb{Z}$. Here the skyrmion number gives the topologically distinct ways that the unit vector field $\boldsymbol{n}$ wraps around the sphere formed by the coordinate space.

A skyrmion number alone is not enough to uniquely determine the spin texture. Figures 11(a) and 11(b) show the vector field of the Bloch type [181] and the Néel type [182] of skyrmions with $N_{sk} = 1$, respectively. To characterize the difference between different types of skyrmions, one can write the unit vector field as $\boldsymbol{n} = (\cos\alpha(\varphi)\sin\beta(r), \sin\alpha(\varphi)\sin\beta(r), \cos\beta(r))$, where $r = \sqrt{x^2 + y^2}$ and $\varphi$ is the polar angle of the 2D coordinate space. Then

$$N_{sk} = \frac{1}{2}\cos\beta(r)\big|_{r=0}^{r=r_\sigma} \frac{1}{2\pi}\alpha(\varphi)\big|_{\varphi=0}^{\varphi=2\pi} = p \cdot m, \tag{2}$$

where $r_\sigma$ denotes the edge of the skyrmion, $p = \frac{1}{2}\cos\beta(r)\big|_{r=0}^{r=r_\sigma}$ represents the polarity, and $m = \frac{1}{2\pi}\alpha(\varphi)\big|_{\varphi=0}^{\varphi=2\pi}$ is the vorticity. For instance, $p = 1$ and $m = 1$ for both Figs. 11(a) and 11(b) since the unit vector pointing downward at the center while upward at the edge and the in-plane winding is 1. Thus an additional phase $\gamma$ with $\alpha(\varphi) = m\varphi + \gamma$ should further be introduced to distinguish the difference between the Bloch type and Néel type of skyrmions. Here $\gamma = -\pi/2$ in for the Bloch type skyrmion in Fig. 11(a) and $\gamma = 0$ for the Néel type skyrmion in Fig. 11(b).

There are many generalized forms of the quasiparticles that enrich the Skyrmion family [183]-[185]. For example, two skyrmions with opposite polarities can form a skyrmionium with zero skyrmion number $N_{sk}$ while still having a nontrivial local spin texture distribution. Another important family generalization of topological quasiparticles is meron and antimeron, with topological charge $+\frac{1}{2}$ and $-\frac{1}{2}$, respectively. Meron (antimeron) might present in thin-film magnetic materials when the material's response is largest under an in-plane external magnetic field. The relation between a skyrmion and a meron (antimeron) is illustrated in Fig. 11(c). The unit vector at the center points up or down, and those at the periphery align in the plane, exhibiting nontrivial windings. Generalizations of merons (antimerons) such as those with higher topological charges or combinations of two with the opposite charges (similar as the skyrmionium) can be constructed following similar rules.

Here our review is restricted to 2D systems, and thus other skyrmion-related concepts such as flying skyrmions or optical hopfions are not introduced here. Since skyrmion-related concepts try to build maps between a unit vector field $\boldsymbol{n}$ and a 2D coordinate space. A simple way of classifying them is through how we construct $\boldsymbol{n}$ and the 2D coordinate space. Electromagnetic waves are three-component complex vectors. There are, in general, three approaches to defining a unit vector field with electromagnetic waves. The first approach is to construct standing waves and then use the resulting real electric or magnetic fields as the vector fields (also dumbed as field skyrmions). The second one is to use the spin vectors of the evanescent electromagnetic fields [186]-[193]. The third approach uses the fact that the polarization of light in free space maps onto the Poincaré sphere. Thus the polarization distribution forms a unit vector field with the vector pointing from the origin to the corresponding locations on the Poincaré sphere. For the 2D coordinate systems, two natural choices are the real space and the momentum space.

We start with the optical field skyrmions wherein the electric (magnetic) field plays the role of the vector field $\boldsymbol{n}$. Surface plasmon polaritons (SPPs) on 2D surfaces are good candidates for studying the optical analog of skyrmions. Indeed, optical field skyrmion lattice was first observed within SPPs [194]. The spin vector of evanescent electromagnetic fields (**S**) can also be chosen as the vector field. The spin vector of light is defined as $\mathbf{S} = \text{Im}(\varepsilon \mathbf{E} * \times \mathbf{E} + \mu \mathbf{H} * \times \mathbf{H})/4\omega$, where **E** and **H** are the electric and magnetic fields, ω is the frequency, and ε and μ are the permittivity and permeability, respectively. By tightly focusing a vortex beam (with a nonzero orbital angular momentum), the longitudinal component (parallel to the propagating direction) of the spin vector $S_z$ can be introduced and a skyrmion-like texture is formed in the focal plane near the beam axis [195]. Here the boundary of skyrmion-like texture is not well-defined. Spin-skyrmion lattices and spin-meron lattices can also be constructed through spin-orbital coupling in the vortex beams by adequately choosing the symmetry of the optical fields [196]-[201].

Besides the electric field and spin of evanescent electromagnetic fields, the polarization of light is another choice for the unit vector field $\boldsymbol{n}$. Spatial polarization distribution, say in a vector beam, can simulate the optical skyrmions [202]. Such a kind of vector beam has been proposed before in the name of full Poincaré beams, and its relation to skyrmion physics have only be revealed recently [203]. Going one step further, the unit vector field $\boldsymbol{n}$ can also be the pseudospin of a two-level system. Such a scheme has been used to construct pseudospin skyrmion in a nonlinear media [198].

The optical skyrmions discussed above are mostly defined in the real space. The momentum space (reciprocal space) of a 2D system can also play the role of a coordinate system. The spin texture of electronic systems in the momentum space has already been investigated in topological insulators [204][205]. In photonics, one can project the eigenmodes onto the free propagating waves, which then define the polarization of the eigenmodes. The variation of eigenmodes polarization is related to the Berry phase in the momentum space. A 2D photonic crystal slab with hexagonal lattice is known to process Dirac cone at the K point and K' point in the reciprocal space. Slightly breaking the inversion symmetry induces a π Berry phase localized around the valley, and this π Berry phase further causes a meron (antimeron) of polarization distribution [206]. Such a photonic crystal slab can also be used to generate light bullets carrying meron spin texture [207].

## 3. Active topological metasurface

Due to the exotic functionalities including unidirectional light propagation and immunity to disorder or defect, topological metasurfaces have attracted extensive attention in both fundamental research and practical applications. However, most of current topological metasurfaces are passive, and it primarily focused on the development of static photonic systems to realize a specific photonic topological phenomenon or functionality, and hence their nonlinearity and reconfigurability are limited. In this section, we review the cutting-edge studies on active topological metasurfaces, which is split into two parts: nonlinear topological metasurfaces and reconfigurable topological metasurfaces.

### 3.1 Nonlinear topological metasurface

Recent works have shown that combining topological metasurface and nonlinear optics can produce a variety of promising applications and important phenomena, including topologically protected frequency converters, high-resolution sensor, topological solitons and modulators, topological lasers, as well as nonlinearity-induced topological phase transitions. These developments not only demonstrate that the active topological metasurface can result in some emerging phenomena and functionalities not presented in the passive regimes, but also suggest that nonlinearity can offer promising applications for novel active devices. As listed in Table 1, we review some typical nonlinear effects and the corresponding applications exhibited in nonlinear topological metasurfaces, including the frequency conversion, parametric amplification, Kerr effect, etc.

Table 1: Typical nonlinear effects in active topological metasurface

| Effect | Platform | Application | Medium | Frequency | Results | Refs |
|---|---|---|---|---|---|---|
| SHG | HOTI | FC | dielectric | ~192 THz (1.55 um) | Sim | [208] |
| | BIC | FC | GaSe | ~225 THz (1.33 um) | Exp | [255] |
| | QVH | Sensing | dielectric | ~0.26 c/a | Sim | [218] |
| | HOTI | FC | Te | 28.4 THz | Sim | [253] |
| | QSH | FC | AlGaAs | ~193 THz | Sim | [254] |
| | BIC | FC | LiNBO$_3$ | ~192 THz (1.55 um) | Exp | [256] |
| | BIC | FC | Si | ~192 THz (1.55 um) | Exp | [217] |
| THG | QSH | Imaging | Si | ~192 THz (1.55 um) | Exp | [219] |
| | BIC | FC | Si | ~212 THz (1.41 um) | Exp | [220] |
| | QH | FC | dielectric | ~0.2 c/a | Sim | [252] |
| | HOTI | Imaging | Si/Al | ~185 THz (1.62 um) | Exp | [221] |
| FWM | QH | QS | Si | ~192 THz (1.55 um) | Exp | [223][251] |
| | QH | Amplifier | graphene | ~13 THz | Sim | [222] |
| | QH | FCB | Si | ~192 THz (1.55 um) | Sim | [261] |
| | QVH | PE | Si | ~197 THz (1.52 um) | Sim | [262] |
| | Floquet | Emitter | Si | ~192 THz (1.55 um) | Exp | [108] |
| HHG | BIC | FC | Si | ~78.1 THz (3.81 um) | Exp | [225] |
| | BIC | OAM | Si | ~191 THz (1.57 um) | Sim | [224] |
| Kerr | Floquet | Modulator | Silica | ~375 THz (0.8 um) | Exp | [259] |
| | Floquet | Soliton | B$_2$O$_3$/Silica | ~291 THz (1.03 um) | Exp | [257][258] |
| | HOTI | Soliton | Silica | ~375 THz (0.8 um) | Exp | [226] |
| | QSVH | Modulator | Si$_3$N$_4$ | ~220 THz (1.36 um) | Sim | [260] |
| Other | QH | Laser | InGaAsP/YIG | ~196 THz (1.53 um) | Exp | [227] |
| | BIC | Laser | InGaAsP | ~192 THz (1.56 um) | Exp | [247][248] |
| | BIC | Laser | GaAs | ~361 THz (0.83 um) | Exp | [249] |
| | QH | Laser | InGaAsP | ~192 THz (1.55 um) | Exp | [229] |
| | QVH | Laser | InGaAsP | ~197 THz (1.52 um) | Sim | [236] |
| | QVH | Laser | GaAs/Al$_{0.15}$Ga$_{0.85}$As | ~3.1 THz | Exp | [235] |
| | HOTI | Laser | InGaAs | ~178 THz (1.68 um) | Exp | [239][240] |
| | BIC | Laser | MAPbBr$_3$ | 543 THz (0.552 um) | Exp | [263] |
| | QSH | Laser | InGaAsP | ~194 THz (1.54 um) | Exp | [230][234] |
| | QSH | Laser | InGaAs/Al$_{0.25}$Ga$_{0.75}$As | ~316 THz (0.95 um) | Exp | [231] |
| | BIC | Laser | Si$_3$N$_4$/IR-792 | ~341 THz (0.88 um) | Exp | [264] |

Notes: Non-Hermitian system (nHs); Quantum source (QS); Photon entanglement (PE); Harmonic generation (HG); Frequency converter (FC); Frequency comb (FCB); Lithium Niobate (LN or LiNBO$_3$); Optical vortex beams (OVB); Orbital angular momentum (OAM); Higher-order topological insulators (HOTIs); Silica and boron trioxide (B$_2$O$_3$); Quantum spin-valley Hall (QSVH) effect; anomalous quantum-Hall (aQH). Sim: Simulation; Exp: Experiment.

### 3.1.1 Frequency conversion

Frequency conversion is a nonlinear optical process where the light interacts with a nonlinear material leading to a generation of new photons with different frequencies. Generally, typical nonlinear effects in frequency conversion include second-harmonic generation, third-harmonic generation, four-wave mixing, high harmonic generation (HHG), Kerr effect, etc. Since the intrinsic nonlinear susceptibility of most nature materials is very weak, it is a critical mission to find a powerful approach to enhance the nonlinear frequency conversion. The field distribution in topological metasurface can be remarkably enhanced in the system edge or corner, thus the active topological metasurface is a promising platform to implement highly efficient frequency conversion and explore novel functionalities.

Second-harmonic generation (SHG), also well known as frequency doubling, is generally characterized by the second-order nonlinear susceptibility of a nonlinear material, where two photons with the same frequency are absorbed to generate a new photon with twice the frequency of the pump photons. Recently, a high-order topological metasurface to use second-order topological corner modes for more efficient second harmonic generation and control has been explored [208]. As shown in Fig.12(a), two topological band gaps of a dielectric photonic crystal are optimized to support corner states that could be frequency matched to realize high efficiency second harmonic generation through the mechanism of double resonance.

High-Q resonances based on BICs have shown tremendous improvement in nonlinear conversion efficiency[141][209]-[216]. In a subwavelength dielectric nanoparticle, two coupled resonances can evolve into a quasi-BIC. The high-Q mode has been theoretically predicted and experimentally demonstrated that nonlinear conversion efficiency in harmonic generation can be enhanced by more than two orders of magnitude[141][211][212]. The design of nonlinear metasurfaces has been inspired to achieve giant nonlinear effects through BICs. Liu et al. have introduced structure perturbation to convert a symmetry-protected BIC into a quasi-BIC for excitation convenience[209]. The Q factor and resonance wavelength are tunable by engineering the structure perturbation and sample size. Tremendous third-harmonic generation and second-harmonic generation have been experimentally observed in the nonlinear metasurface. Continuous-wave second harmonic generation enhanced by BICs has also been demonstrated by another group by controlling the asymmetric parameter of metasurfaces[214]. Moreover, an array of slotted nanocubes is also designed to obtain remarkable SHG from the centrosymmetric silicon, by taking advantage of strengthened electric field distribution, enlarged surface second-order nonlinearity, and the resonance-induced enhancement by BIC [217], as shown in Fig.12(b). The corresponding experimental results show that the SHG from the slotted nanocube array is improved by more than two orders of magnitude, compared with that from the array of silicon nanocubes without air-slots.

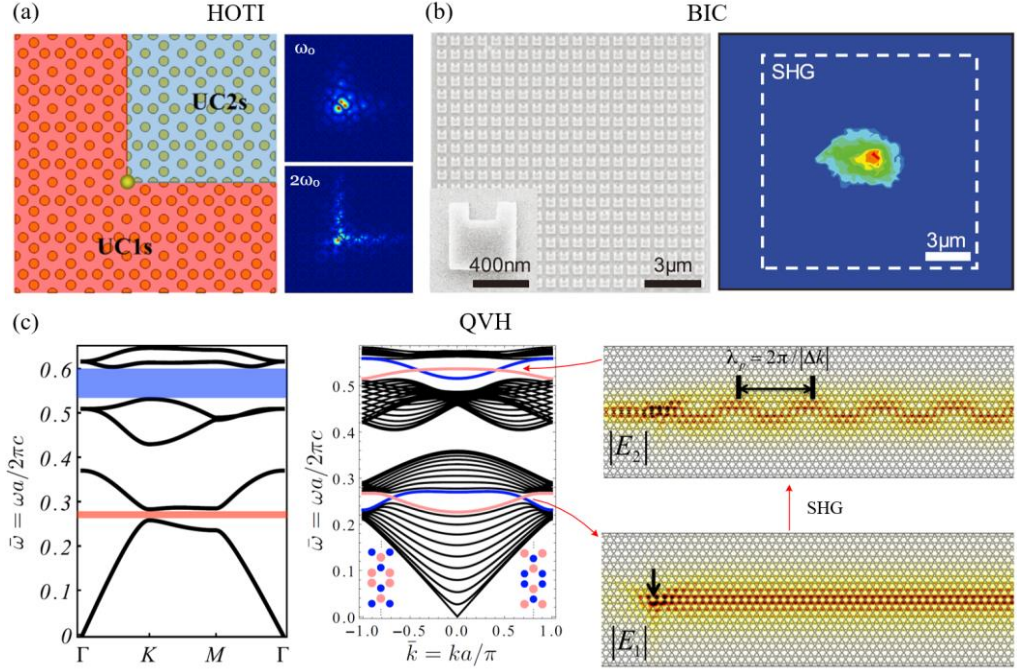

**Fig. 12.** Second-harmonic generation (SHG) in active topological metasurface. (a) SHG mediated by two corner modes which reside within two different topological bandgap and could be frequency-matched to greatly boost the harmonic conversion efficiency by the mechanism of double resonance [208]. (b) Spatial mapping of the SHG in a fabricated slotted nanocube array [217]. (c) Band diagram and simulated field intensities of the fundamental ($E_1$) and second-harmonic ($E_2$) waves in a dielectric metasurface [218].

Based on analog quantum valley Hall effect, the topological metasurface is also a good platform to explore novel phenomena in SHG. For example, an all-dielectric topological metasurface, which has double topological valley-Hall kink modes, was proposed recently [218]. By gapping out the corresponding Dirac points, two topological frequency band gaps can be created around a pair of frequencies, namely the fundamental and second-harmonic frequencies. The corresponding numerical results prove that valley-Hall kink modes along a kink-type domain wall interface can be generated within the two frequency band gaps, where tunable, bidirectional phase-matched SHG via nonlinear interaction of the valley-Hall kink modes can be achieved, as demonstrated by the SHG field distributions in Fig.12(c).

Third-harmonic generation (THG), also named as frequency tripling, is generally characterized by the third-order nonlinear susceptibility of a nonlinear material, where three photons with the same frequency are absorbed to generate a new photon with thrice the frequency of the pump photons. As depicted in Fig. 13(a), based on the higher-order topological effect, a silicon topological metasurface, which can support topologically protected helical edge states, has been experimentally studied [219], recently. Due to enhancement boosted by multipolar Mie resonances of silicon nanoparticles, a strong THG is measured. Moreover, the independent high-contrast imaging of either bulk modes or spin-momentum-locked edge states was explored under different pump-beam wavelengths. They also demonstrate the pseudospin-dependent unidirectional waveguiding of the edge states bypassing sharp corners.

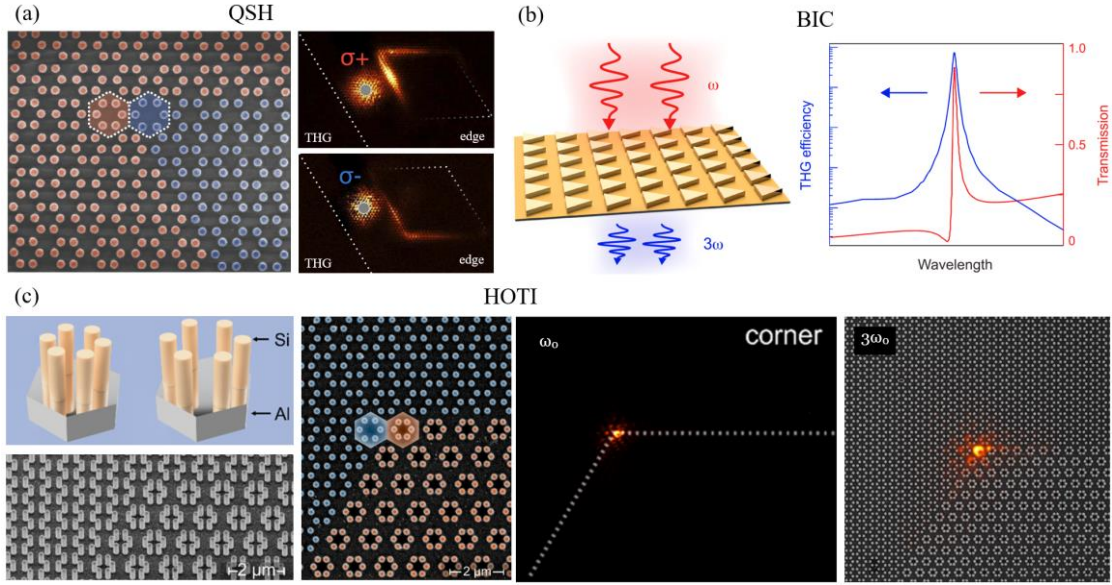

**Fig. 13.** Third-harmonic generation (THG) in active topological metasurface. (a) THG in QSH topological metasurface consisting of silicon pillars arranged into hexagon clusters [219]. (b) THG in a nonlinear and asymmetric metasurface governed by BIC [220]. (c) THG enhanced by a topologically protected edge mode in high-order topological metasurfaces [221].

In addition to the quantum spin Hall effect, based on the bound states in the continuum, a silicon-based metasurface composed of meta-atoms with broken in-plane symmetry has been designed to tailor the THG efficiency by engineering the degree of the unit cell asymmetry [220], as shown in Fig. 13(b). Furthermore, the effect of radiative and nonradiative losses on the nonlinear conversion efficiency is discussed based on the concept of the critical coupling of light to the metasurface resonances. By tuning the metasurface parameters to the regime of the critical coupling, the maximum efficiency of the frequency conversion in topological metasurfaces can be achieved, when the contributions of radiative and nonradiative loss mechanisms coincide.

Moreover, based on the concept of high-order topological insulators, a hybrid metal-dielectric topological metasurface with symmetric honeycomb lattices is recently proposed to enhance the intensity and robustness of THG [221]. Since the topological edge and corner states can not only enhance the light intensity but also offer a robust protection of light against the disorder and defect perturbations, a significantly enhanced THG has been experimentally observed. As shown in Fig. 13(c), such a remarkable improvement can be used to develop a nonlinear imaging technique, which enables a high-resolution and background-free photograph of the states.

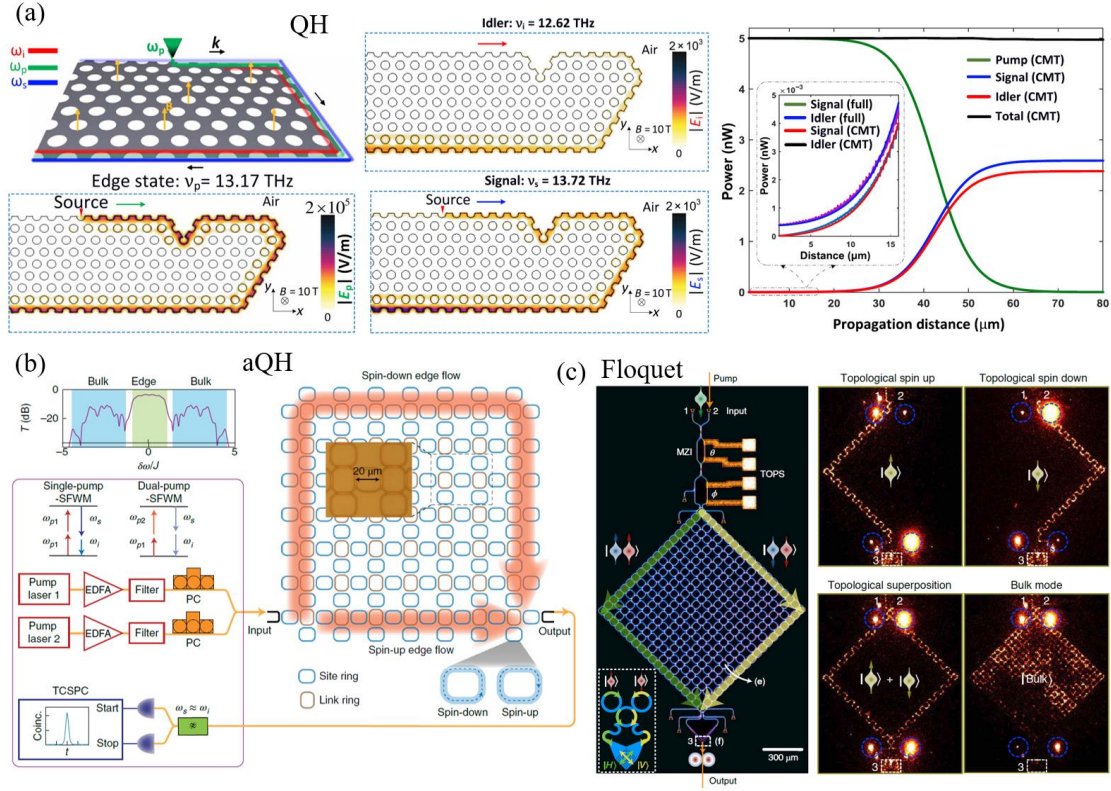

**Fig. 14.** Four-wave mixing (FWM) in active topological metasurface. (a) FWM of topologically protected one-way edge plasmons in a graphene QH topological metasurface [222]. (b) Generation of indistinguishable photon pairs via spontaneous FWM in an anomalous QH topological metasurface [223]. (c) Entangled photons emerge and flow at a pair of edge modes in a silicon anomalous Floquet topological metasurface [108].

Four-wave mixing (FWM) is an intermodulation process in nonlinear optics, and it is characterized by the third-order nonlinear susceptibility of a nonlinear material as similar as the case in THG. However, the difference is that the frequency of the three pump photons are not the same. In FWM nonlinear interaction, two or three frequencies generate one or two new frequencies. The confined edge or corner light modes can significantly enhance the strength of FWM nonlinear process. For example, a plasmonic QH topological metasurface consisting of a periodic array of nanoholes in a graphene sheet was proposed to study the topological protection and enhancement of FWM interactions [222]. As demonstrated in Fig. 14(a), the QH effect is realized by breaking the time reversal symmetry via externally applying a static magnetic field. Due to the significant nonlinearity enhancement and large life time of graphene plasmons along the topological edge states, a net gain of FWM interaction can be achieved with an ultralow pump power of less than 10 nW, which is a consequence of the unusually large effective nonlinear edge-waveguide coefficient $\simeq 1.1\times10^{13}$ $W^{-1}m^{-1}$. It is more than 10 orders of magnitude larger than that of commonly used silicon photonic crystal.

In addition to theoretical studies, some important experimental results are reported recently. For instance, an anomalous QH topological metasurface consisting of coupled ring resonators is reported to generate indistinguishable photon pairs via spontaneous FWM [223]. Due to the linear dispersion of the topological edge states, a phase-matched generation of photon pairs throughout the edge band can be easily achieved. Thus, by tuning the input pump frequencies in the edge band, the spectral-temporal bandwidth of photon pairs can be engineered. Since the anomalous QH topological metasurface is a time-reversal symmetric system, it supports two spin-locked topological edge states, as shown in Fig. 14(b). Such counter-propagating topological edge states can be used to generate path-entanglement and split the indistinguishable photon pairs. Furthermore, entangled states are generally sensitive to the disorder or defect in optical devices, whereas topological states are naturally robust against fabrication perturbations. Thus, it has great potential for generating topological protected entanglement via FWM. Recently, based on the complementary metal-oxide-semiconductor (CMOS) fabrication technique, a silicon anomalous Floquet topological metasurface has been proposed to experimentally demonstrate the robustness of a topological photonic entanglement against certain imperfections [108]. As depicted in Fig. 14(c), the anomalous Floquet topological metasurface consists of a square lattice of coupled silicon ring resonators, and support two counter-propagating edge states with identical spectral distribution. To demonstrate the robustness of a topological photonic entanglement, some imperfections has been introduced by adding or removing a single ring or an entire unit cell in the ring lattice structure. The experimentally results prove that the topological edge states can be used to protect the photonic entanglement in the presence of structure defects.

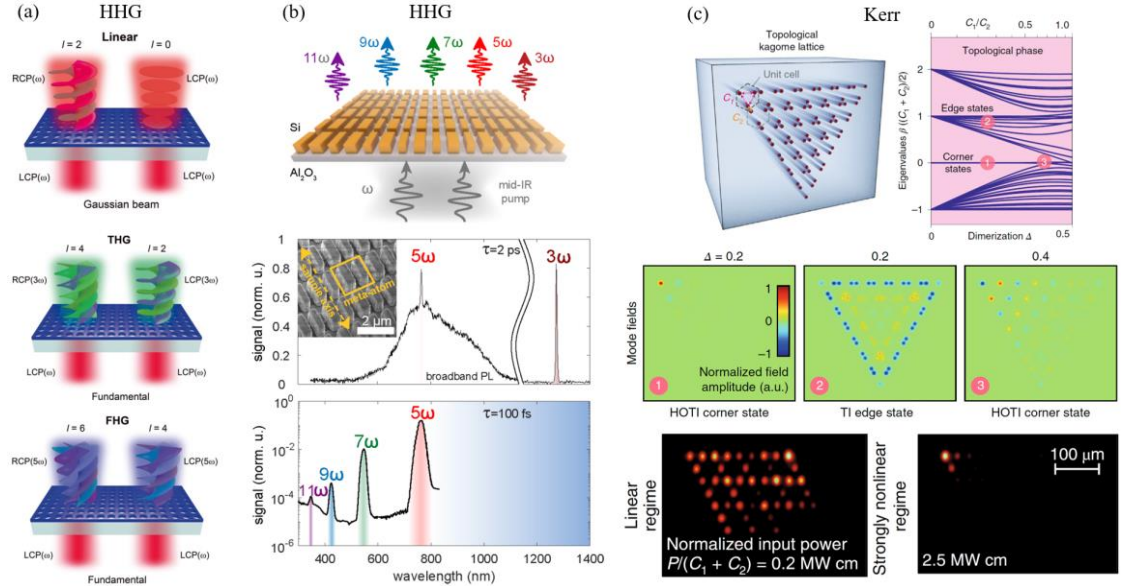

**Fig. 15.** High harmonic generation (HHG) and Kerr effects in active topological metasurface. (a) High-harmonic optical vortex generation in a symmetric BIC topological metasurface [224]; (b) 3rd to 11th optical harmonics generated in a nonlinear and asymmetric BIC topological metasurface [225]; (c) Power-dependent corner states and solitons in high-order topological metasurfaces [226].

High harmonic generation (HHG) is a nonlinear interaction between an intense light and nonlinear materials to generate new photons with high harmonic frequencies, which are generally above fifth harmonic. Recent studies find that topological metasurface is a particularly fertile platform for the study of HHG. For example, an amorphous silicon BIC topological metasurface has been designed to generate the third- and fifth-harmonic optical vortex [224]. In the linear regime, when the topological metasurface is illuminated by a circularly polarized Gaussian beam, it only generates the opposite-handedness transmitted beam with a certain topological charge. But in the nonlinear regime, under the same excitation condition, the same topological metasurface can generate the THG and the FHG signals, which enables the harmonic optical-vortex generation of both handedness, as shown in Fig. 15(a). Moreover, the simulated results show that the topological charges in the nonlinear case are related to the order of the harmonic signals. In addition to the symmetric BIC effect, an asymmetric BIC topological metasurface has also been proposed to study the HHG [225]. To support bound states in the continuum, a pair of rectangular silicon bars in each unit cell is optimized to break the in-plane symmetry, and an asymmetry parameter is defined, as shown in Fig. 15(b). The corresponding results prove that the radiative and quality factor of the quasi-BIC are highly dependent on the asymmetry parameter. By illuminating intense laser pulses on the asymmetry topological metasurface, the generation of optical harmonics up to the 11th order can be experimentally measured. The experimental results also prove that the HHG is highly dependent on the pump polarization and pulse duration.

Different from aforementioned frequency conversions, Kerr effect is a special nonlinear process, as there is only a nonlinear change in the refractive index of a material, without any generation of new frequencies. In fact, Kerr effect includes the electro-optic Kerr effect, also called direct-current (DC) Kerr effect and the optical Kerr effect, also well-known as alternating-current (AC) Kerr effect. It is generally characterized by the real part of the third-order susceptibility, whose imaginary part indicates a nonlinear absorption, namely optical saturable absorption. Kerr effect in active topological metasurface has promising potential applications for all-optical modulator, optical soliton, etc. For instance, a high-order topological metasurface has been designed to experimentally demonstrate the nonlinear corner states as well as the formation of solitons [226]. As shown in Fig. 15(c), the high-order topological metasurface is constructed by a kagome lattice of fused silica waveguides. When the strength of the intra-cell bond $C_1$ exceeds that of the inter-cell bond $C_2$, namely ($C_1 > C_2$), the light modes are trivial. But, when $C_1 < C_2$, there are some non-trivial states, also called topological states, including topological edge and corner states, which could be verified by corresponding field distributions. To study the nonlinear Kerr effect, an intense ultrashort laser with enough peak powers is used to illuminate the corner sites of the topological metasurface to elicit a nonlinear Kerr response. In the linear regime, a linear light excitation (0.2 MW cm) penetrates deeply into the topological metasurface. By increasing the input power to 2.5 MW cm, due to the nonlinear Kerr response in the topological metasurface, the light mode is confined to the corner to form a tightly localized soliton. These findings pave the avenue towards the development of nonlinear compact devices.

### 3.1.2 Topological laser

Lasers, as the most representative active devices, have been extensively studied in different kinds of photonic cavities. However, those photonic cavities are usually sensitive to defects, disorders, and fabrication imperfections and suffer from device-to-device performance, which imposes a fundamental constraint on the applications of lasers. Lasers built on topologically robust cavities have, therefore, been highly desired ever since the invention of the lasers.

The first experimental demonstration of topological lasers in QH topological metasurface has been reported in [227], where the nonreciprocal single-mode lasing from topological cavities of arbitrary geometries was designed at room temperature and at telecommunication wavelengths. The nonreciprocal topological laser was implemented by two photonic crystals with distinct topological invariants. Despite the narrow magneto-optically induced photonic bandgap, robust topological cavity mode in arbitrary-shaped cavity and topological lasing can be observed with a significant isolation ratio as large as 11.3 dB from the oppositely propagating mode. However, since the magneto-optic effect is very weak at optical frequencies, only a narrow topological band gap is produced. To enlarge the topological band gap, another all-dielectric topological metasurface based on artificial magnetic fields is theoretically proposed and experimentally demonstrated [228][229], as shown in Fig. 16(a). It possesses a sizable topological band gap and operates under optically pumped and at cryogenic temperatures.

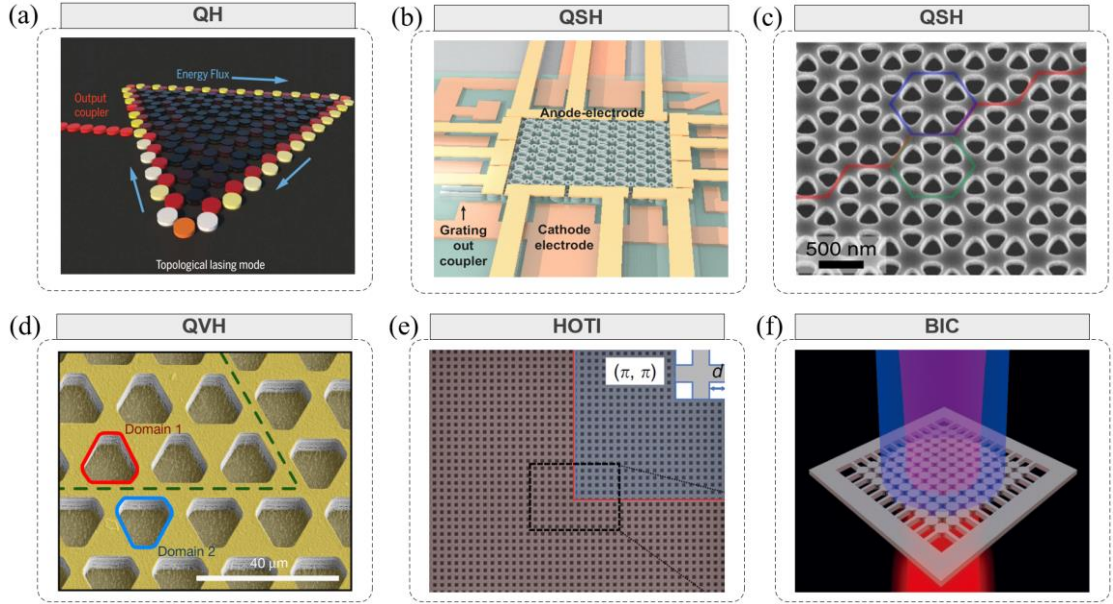

**Fig. 16.** Topological laser based on active topological metasurface. (a) Topological laser in a QH topological metasurface consisting of honeycomb lattice of coupled ring resonators [228]. (b) Electrically pumped topological laser in a QSH topological metasurface comprising of square lattice of ring cavities and link resonators [234]. (c) Topological bulk laser in a QSH topological metasurface based on band-inversion-induced reflection [230]. (d) Electrically pumped topological laser based on QVH effect operating at terahertz frequencies [235]. (e) Low-threshold topological laser in a second-order topological metasurface [239]. (f) Lasing improved by BICs [247].

The applications of QSHE have been extended to construct topological cavity laser by using the topological edge states with spin-momentum locking property. Shao et al [230] proposed a topological bulk laser, where band-inversion, because of inducing SOC in QSHE, is used to realize a total reflection cavity with the mode-mismatch between dipole and quadrupole. Dikopoltsev et al. [231] arrayed the vertical-cavity surface-emitting laser (VCSEL), in which each emitter acts as a single laser. While they are spatially coherent interference, emitters on the topological edge are injection-locked, and lase vertically and coherently. As demonstrated in [232], based on silicon-on-insulator platform, a topological bandgap emerges by applying the generalized Kekulé modulations, which experimentally realizes a Dirac-vortex topological cavity with hosts higher yield, wider tuning range, narrower linewidth, and greater output power. In [233], the Dirac-vortex topological cavity supports optimal single-mode selection in a large area, and can be used to realize a high-power narrow-beam topological-cavity surface-emitting at the most important telecommunication and eye-safe wavelength (1550 nm).

More recently, an electrically pumped topological laser under room temperature was developed [234], as shown in Fig. 16(b). This topological metasurface consisted of a periodic array of resonators coupled through an aperiodic set of auxiliary link structures to mimic the quantum spin Hall effect, thus it can generate topologically protected lasing and exhibit single frequency emission. Apart from lasing to topological edge/interface modes, topological properties based on bulk states could also be used to realize topological lasing. In such a topological bulk laser [230], the topological metasurface shows a topological band inversion around the Γ point between its interior and cladding area. As shown in Fig.

16(c), due to the band inversion, the wavefunctions of bulk states in topologically trivial and nontrivial areas have opposite parities. As a result, in-plane light waves in the trivial region cannot propagate into the nontrivial region, and they are reflected at interface forming an effective cavity feed-back. Furthermore, the band-inversion-induced reflection only occurs around the Γ point, which provides a novel lasing mode selection mechanism and enables directional lasing emission of cavity modes.

It has also been demonstrated that valley-Hall PTIs can be utilized to construct topologically robust cavities by creating a domain wall in a closed form [235]-[238]. For example, the topological cavity can be applied to devise the electrically pumped THz quantum cascade lasers (QCL), which are widely used in communication, imaging, and sensing at the THz domain [235]. As shown in Fig. 16(d), the laser cavity is a triangle loop supporting robust valley-kink states, and the yellow region is electrically pumped. With an electrical injection, the running-wave mode in the cavity is lasing and distributes uniformly around the domain wall even in the presence of defects and sharp corners. The robust laser based on the valley kink states was also realized at optical frequencies. Moreover, room-temperature lasing with a narrow spectrum, high coherence, and threshold behaviour was experimentally observed in this laser. Interestingly, the emitted beam exhibits a singularity encoded by a triade cavity mode that locates at the three corners of the cavity [238].

Due to the small footprint (small mode volume and high quality factor), second-order topological corner state has also been exploited for topological lasing [238]-[241]. In [239], the first experimental observation of a topological nanolaser in a 2D topological PhC nanocavity, exploiting the Wannier-type 0D corner state at the nanoscale, was demonstrated in Fig. 16(e). In [241], topological lasing was observed at all hierarchical eigenstates, i.e., 2D bulk, 1D edge, and 0D corner states in a 2D PhC platform with a square area of a topologically nontrivial PhC structure surrounded by a topologically trivial counterpart. Furthermore, multiple corner states and their interactions have also been observed, e.g., in [238], triade mode lasing, in which three corner modes are coupled to each other and exhibit lasing action, was observed at the nanoscale and room temperature in a valley-Hall nanophotonic cavity embedded into 2D topological lattices with semiconductor quantum wells as a gain medium.

BICs have manifested superiority in boosting lasing performance by reducing the lasing threshold and miniaturizing the size of lasers. Although many surface-emitting lasers[242]-[246] are based on symmetry-protected BICs, the first announcement of BIC lasing is demonstrated by Kodigala[247] et al. In a square lattice formed by InGaAsP multiple quantum wells cylindrical nanoresonators, as shown in Fig. 16(f), off-Γ BICs are tuned to the degenerate state at the Γ point for optimizing. The lasing threshold is experimentally confirmed lowest at BICs. Furthermore, lasing action persists even though the cavity has been miniaturized to 8-by-8 nanoresonators. Hwang et al. have experimentally demonstrated an ultra-low threshold laser, where the scattering loss in a finite-size cavity has been further suppressed by merging multiple BICs[248]. Ha et al. have experimentally realized directional lasing empowered by BICs in dielectric nanoantenna arrays[249]. The system has a symmetry-protected BIC to improve light confinement. By adjusting one of the periods, a diffraction order is allowed to achieve directional lasing. Bahari et al. have exploited polarization vortexes around BICs to realize vortex lasers[250].

### 3.2 Reconfigurable topological metasurface
Generally, there are two major constraint of the unreconfigurable topological metasurface: (1) The space utilization ratio, namely the ratio of edge region to the bulk region, of most passive topological metasurfaces, is very low. Specifically, since the topological modes (such as edge mode, corner mode) exist only at the boundary or corner, most region (such as bulk region) of topological metasurface is wasted, hindering the high-density opto-electronic integration. Owing to the reconfigurable functionality of active topological metasurface whose light propagation route can be reconfigured on demand inside bulk region, its space utilization ratio can be significantly improved, leading to a high-density photonics routing. (2) The optical performances of passive topological metasurfaces are fixed once the device has been fabricated. However, for many practical applications, the reconfigurability and tunability of light manipulation are essential. For instance, multiple photonic topological functionalities are expected to be achieved in a single but reconfigurable PTI, so that the time and costs associated with the design and fabrication process can be reduced. Currently, the reconfigurable manners of topological metasurfaces include electrical control, optical control, mechanical control and thermal control, as listed in Table 2.

Table 2: Typical reconfigurable manners in active topological metasurface

| Mechanism | Platform | Medium | Frequency | Tuning Time | Result | Refs |
|---|---|---|---|---|---|---|
| Electrical | QVH | Graphene | 28.3 THz | / | Sim | [267] |
| | QSH | LC | 0.44c/a | / | Sim | [268] |
| | QVH | BaTiO$_3$ | ~0.38c/a | / | Sim | [281] |
| | QVH | LC | ~14 THz | / | Sim | [282] |
| | QVH | PIN | 7.2 GHz | ~50 ns | Exp | [265] |
| | QH | LC | / | / | Sim | [283] |
| | HOTIs | LC | ~28 THz | / | Sim | [269] |
| | QVH | LC | ~0.35 THz | / | Sim | [270] |
| | Floquet | TS | 0.5 GHz | 2 ns | Exp | [266] |
| Optical | Floquet | AZO | ~230 THz (1.3 um) | / | Sim | [271] |
| | QVH | Si | ~182 THz (1.64 um) | >0.6 ns | Exp | [272] |
| | non-Hermitian | InGaAsP | 202 THz (1.486 um) | / | Exp | [273] |
| | BIC | MAPbBr$_3$ | 543 THz (0.552 um) | ~1.5 ps | Exp | [263] |
| Mechanical | QSHE | metallic | ~20 GHz | ~3s | Exp | [40] |
| | SSH | dielectric | ~428 THz (0.7 um) | / | Sim | [274] |
| | QSHE | Si | 0.46 c/a | / | Sim | [275] |
| | QSHE | dielectric | ~0.74 c/a | / | Sim | [276] |
| Thermal | QSHE | Ge$_2$Sb$_2$Te$_5$ | ~138 THz (2.174 um) | ~150 ns | Exp | [277] |
| | Floquet | TiN | 230 THz (1.305 um) | / | Sim | [278] |
| | HOTIs | Sb$_2$S$_3$ | 194 THz (1.55 um) | ~5 min | Sim | [279] |
| | QVH | VO$_2$ | 0.16 THz | / | Sim | [280] |

Notes: Aluminum-doped zinc oxide: (AZO); Su-Schrieffer-Heeger (SSH) model; Liquid crystal (LC); Higher-order topological insulators (HOTIs); transistor switches (TS); Vanadium(IV) oxide (VO$_2$); Sim: Simulation; Exp: Experiment.

So far, the electrically reconfigurable manners of topological metasurfaces are mainly studied theoretically, and there are few experiment reports, in which the operating frequencies are all in microwave regime. The most explored physical effect in current reconfigurable topological metasurfaces is quantum valley Hall effect.

In microwave regime, it is relatively convenient to integrate active lumped elements (such as PIN and other transistors) into topological metasurface to implement an electrically reconfigurable functionality. For example, You *et al.* have reported the first experiment result of electrically reconfigurable topological devices [265], as shown in Fig. 17(a). The reconfigurable manner of this topological metasurface is implemented by judiciously designing each programmable unit cell, which has six inner arms bridged with one outer arm via a PIN diode. The six outer arms are connected to a field-programmable gate array (FPGA) control network by six metallic via holes. The states of the PIN diode can be dynamically switched between 'on' and 'off', thus to emulate the binary states of '1' and '0'. In this way, the geometry of the domain wall can be actively changed, enabling dynamic control of topological photonic transport paths. Due to the flexible programmability and the innovative use of electric switches, the topological light propagation route can be dynamically changed at nanosecond-level switching time (almost 50 ns). Recently, a faster switching time (2 ns) is achieved in a chip-scale Floquet topological metasurface [266]. As shown in Fig. 17(b), it is consisted of 4×4 helicoidally rotating quasi-electrostatic circulator unit cells. The electrical control is implemented by switching the transistor bridging a shunt capacitor with a four-port unit circulator.

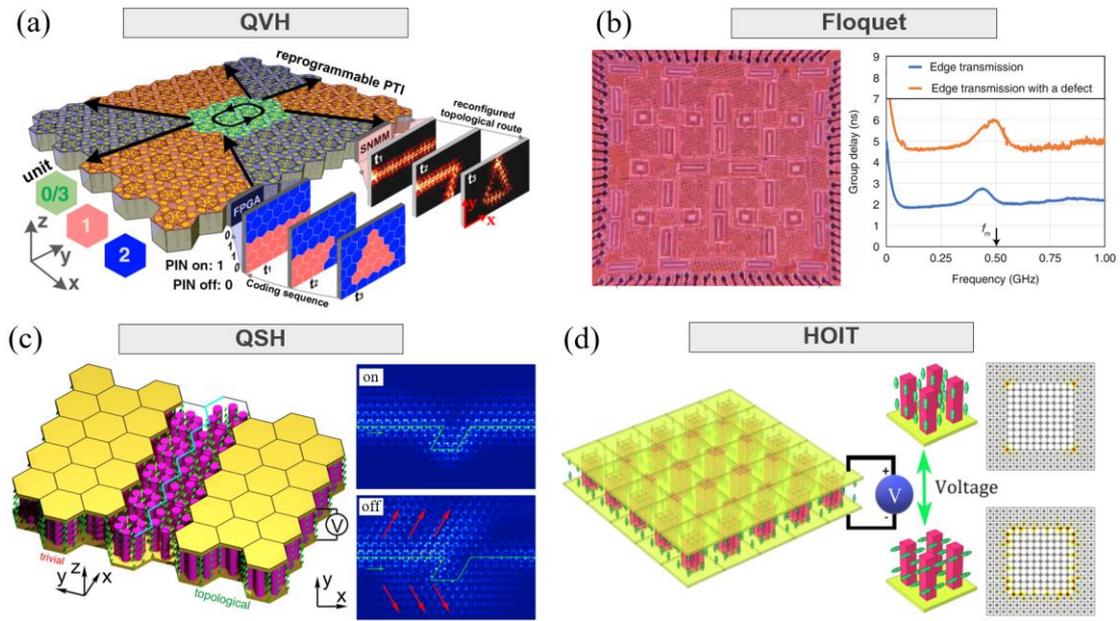

**Fig. 17.** Electrically controlled reconfiguration in topological metasurface. (a) Ultrafast reprogrammable plasmonic topological metasurface based on QVH effect [265]. (b) Chip-scale Floquet topological metasurface based on switched-capacitor networks [266]. (c) Reconfigurable QSH topological metasurface based on liquid crystal [268]. (d) HOTI topological metasurface supporting edge-corner state switching [269].

In addition to the lumped element, some other electro-optical materials (such as graphene, BaTiO3, liquid crystal, etc.) can also be used to implement the electrical control of topological metasurfaces. For example, since the Fermi energy of graphene can be tuned electrically, a mid-infrared topological metasurface in metagated-tuned graphene was proposed to enable rapid switching of topological plasmons via simple electric gating [267]. Meanwhile, the liquid crystal is another popular phase-change material widely used to implement electronically controlling topological metasurface, as the anisotropic permittivity of liquid crystal is tunable under different bias voltages. As shown in Fig. 17(c), based on the quantum spin Hall effect, a topological metasurface consisting of silicon pillars surrounded by a liquid crystal was studied [268]. The topological metasurface is enclosed between conducting electrodes, and it can be switched by applying voltage to the electrodes, as the refractive index of liquid crystal would be changed when the voltage is applied.

Recently, it reports that the corner states can also be electrically controlled by liquid crystal [269]. As shown in Fig. 17(d), a dynamically tunable and reconfigurable topological metasurface is realized in higher-order topological insulators by changing the loading voltage of the liquid crystal. In this reconfigurable topological metasurface, its edge and corner states can be switched at the same frequency. In addition to the HOTIs, a tunable and programmable valley topological metasurface based on nematic liquid crystals has also been reported [270], and its inversion symmetry breaking and topological transition are implemented through electrically controlling the relative permittivity of the LC cells. Moreover, this valley topological metasurface can be discretized to a number of supercells, each of which is coded with '0' or '1' to realize a programmable functionality.

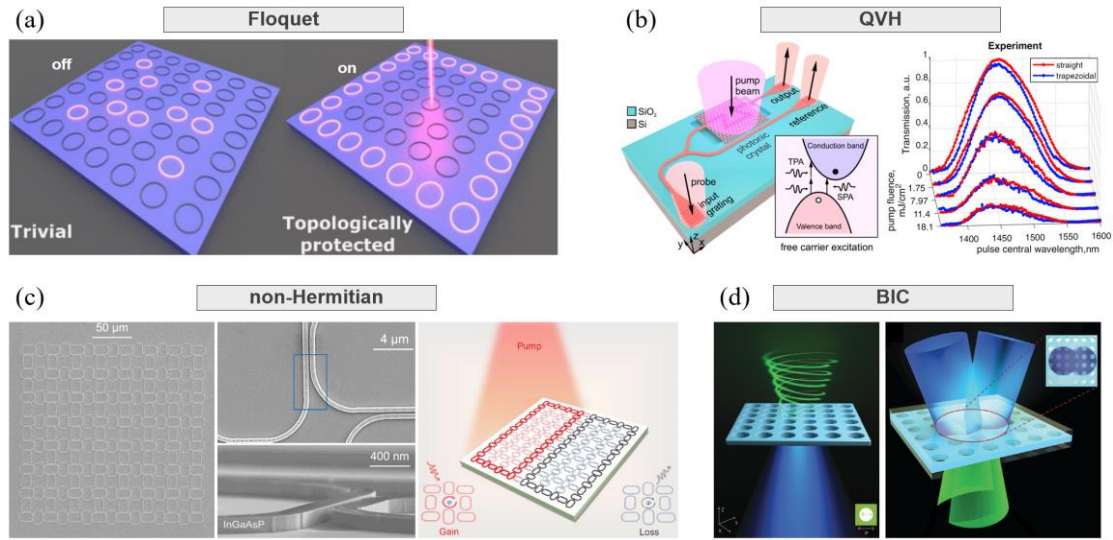

**Fig. 18.** Optically controlled reconfiguration in topological metasurface. (a)All-optical control of topological states in a Floquet topological metasurface [271]. (b)Transmission modulation in a QVH topological metasurface by optically tuning the refractive index of silicon [272]. (c) Optically reconfigured the topological edge states by breaking local non-Hermitian symmetry [273]. (d) Ultrafast all-optical switching between the vortex beam lasing and linearly polarized beam lasing [263].

Optically controlled reconfigurable light propagation routes in topological metasurface were reported in [263][271]-[273]. For instance, based on the tunable transparent conducting oxides (TCO) and standard silicon-on-insulator wave-guide technology, a Floquet topological metasurface has been theoretically developed to achieve ultrafast all-optical control of the topological states [271]. As depicted in Fig. 18(a), the unit cell of proposed topological metasurface consists of four ring resonators and four link couplers, which have been covered with aluminum-doped zinc oxide (AZO). AZO is a typical kind of TCO materials, and its refractive index is sensitive to the pump light. Therefore, an ultrafast control over the mode index of link couplers can be realized by switching on or off the pump light, leading to the all-optical control of topological states in a Floquet topological metasurface. In addition to theoretical results, based on the quantum valley Hall effect, a silicon-on-insulator topological metasurface with an optically controlled reconfigurable functionality has been experimentally studied recently [272]. The tunability is enabled by the free-carrier excitation initiated by the pump beam, leading to a reduction of the real part of the refractive index and increase of the imaginary part of the refractive index in silicon. As shown in Fig. 18(b), the fast refractive index modulation could be used to switch the topological light propagation, and the corresponding switching time is up to the order of nanoseconds.

In addition to silicon, InGaAsP can also be used to construct optically controlled topological metasurface. As an example, a non-Hermitian topological metasurface was proposed recently based on optical modulation [273]. As demonstrated in Fig. 18(c), the topological metasurface consists of coupled microring resonators, and the optical control is realized by illuminating a laser on the InGaAsP metasurface to create gain (pumping region) and loss (no pumping region) domain walls, where the topological states can be observed. In this way, the optical control can be used to actively steer topological light propagation route by projecting a spatial pumping pattern onto the topological metasurface. The direction of vortex lasers is steerable by tuning BICs in the momentum space, as shown in Fig. 18(d). Huang et al. have experimentally demonstrated an ultrafast all-optical switching of vortex microlaser using the topological nature of BICs[263]. Winding polarizations around a symmetry-protected BIC generate vortex emission. When symmetry is reduced under asymmetric excitation, lasing becomes a linearly polarized beam. Ultrafast switching between the vortex beam lasing and linearly polarized beam lasing can be realized by controlling two-beam pumping.

Mechanically controlled reconfigurable electromagnetic pathways in topological metasurface were reported in [40][274]-[276]. Based on the quantum spin Hall effect, a reconfigurable topological metasurface has been experimentally studied between two parallel copper plates [40]. To implement the mechanically controlled reconfigurable functionality, a periodic triangular array of metallic rods with ring collars drill the holes on the parallel copper plates, as illustrated in Fig. 19(a). By mechanically moving the rods up or down, a reconfigurable electromagnetic pathway between two topologically distinct domains can be created on demand.

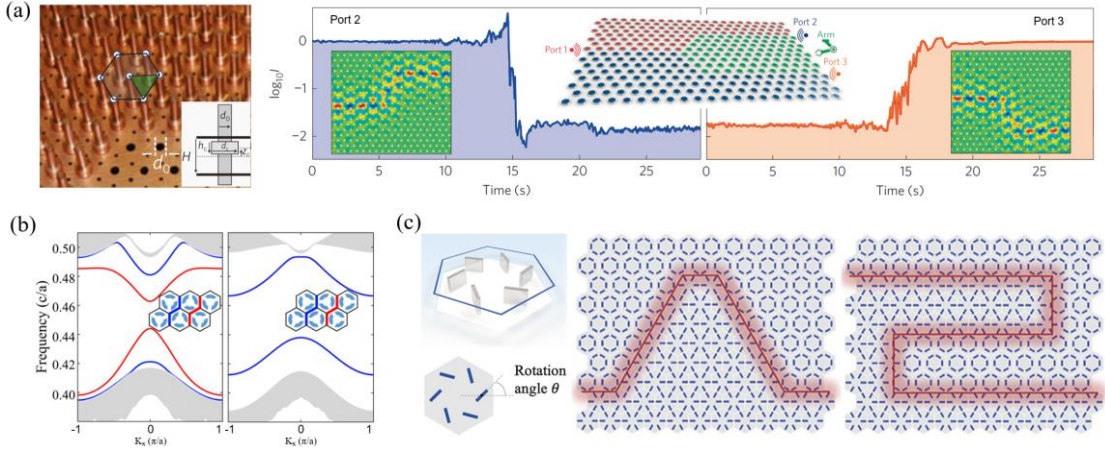

**Fig. 19.** Mechanically controlled reconfiguration in topological metasurface. (a) Robust reconfigurable microwave propagation routes in a QSH topological metasurface [40]. (b) Tunable edge states in a split-ring topological metasurfaces [275]. (c)Reconfigurable topological metasurface based on honeycomb lattice of rotating dielectric cuboids [276].

Split-ring structures can also be used to construct a mechanically controlled reconfigurable topological metasurface [275]. As shown in Fig. 19(b), since the topological band diagrams depend on the rotation angle of the split-ring unit, a tunable topological state can be achieved by mechanically controlling the rotation angle to introduce two topologically distinct phases. Similar mechanically control in reconfigurable topological metasurface has also been implemented theoretically in a honeycomb lattice with dielectric cuboids [276]. Specifically, the unit cell of designed topological metasurface consists six dielectric cuboids with a rotation angle. By mechanically rotating the six cuboids around their own centers, a radial or circular structure in each unit cell can be constructed to open or close the topological bandgap, as demonstrated in Fig. 19(c). As a consequence, the light propagation routes can be mechanically reconfigured in the proposed topological metasurface.

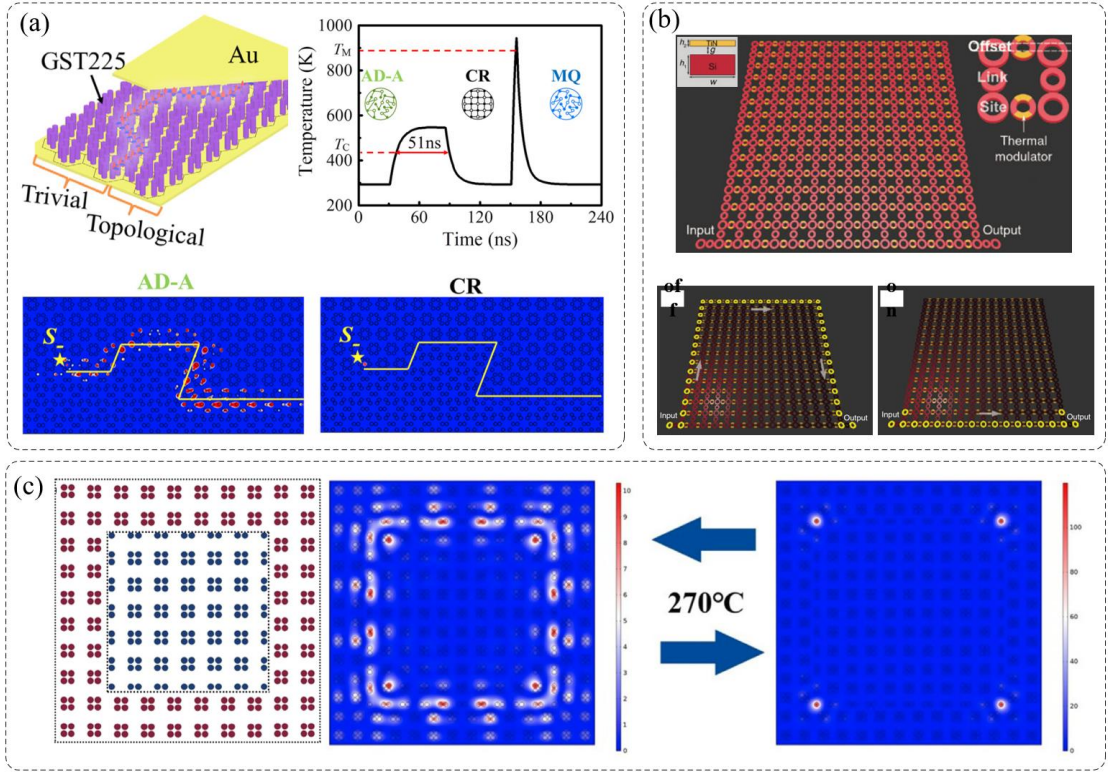

**Fig. 20.** Thermally controlled reconfiguration in topological metasurface. (a) Dynamically reconfigurable topological edge state in a thermally controlled topological metasurface [277]. (b) Thermally controlled topological metasurface based on silicon-on-insulator technology [278]. (c) Thermally controlled edge and corner states in HOTI topological metasurfaces [279].

Thermal control is another important approach to construct a reconfigurable light propagation route in topological metasurface [277]-[280]. Based on the quantum spin Hall effect, a dynamically reconfigurable edge state in topological metasurfaces has been experimentally studied [277]. By changing a thermal phase-change material, namely $Ge_2Sb_2Te_5$, between amorphous and crystalline, the refractive index of $Ge_2Sb_2Te_5$ pillar unit could be thermal modulated, leading to a reconfigurable topological edge state, as shown in Fig. 20(a). It demonstrates that the topological metasurface can fast switch the topological light propagation in 150 ns.

Thermally controlled edge states can also be realized in a Floquet topological metasurface [278]. As demonstrated in Fig. 20(b), its unit cell includes four "site" ring resonators and four "link" ring couplers. The couplers are made of silicon and covered with thermal modulators. Based on the thermo-optic effect in silicon, the accumulated phase of light propagating through the rings can be modified to control the topological edge states. In addition to thermally controlled edge states, a tunable and reconfigurable high-order topological state in topological metasurface with phase change materials, namely $Sb_2S_3$ and $Sb_2Se_3$, has also been thermally controlled [279]. More specifically, an SSH-like square topological metasurface is optimized. As depicted in Fig. 20(c), its outer region consists of the topological trivial unit cells, whereas its inner region is filled with topological nontrivial unit cells. Since the refractive index of phase change materials is sensitive with temperature, the topological edge state and topological corner state at the same frequency can be switched relatively fast.

# 4. Beyond

Exploring and harnessing quantum mechanical effects for future technological applications have attracted ever-increasing attention in recent years. However, in most cases, quantum features of materials are quite fragile and susceptible to their structural as well as environmental changes and require stringent conditions in order for their quantum properties to maintain, e.g., it is well known that entangled states are fragile with respect to decoherence in a noisy environment. This fragility of quantum states is particularly relevant for quantum photonic states in nanophotonic structures where small fabrication imperfections are unavoidable and these imperfections can have a detrimental impact on the quantum information encoded in the photonic states. As the system properties of topological matters only depend on the global properties (topological invariants) of the system, where local defects and imperfections will have no effect on the system properties, the concept of topology could provide a promising route to design highly robust quantum photonic devices with built-in topological protection, whose quantum properties can be much easier to maintain. Thus it is particularly interesting to apply topological concepts to protect the fragility of photonic states in the quantum regime [284][285].

In this section, we discuss how topological photonics can advance the study of quantum optics. We first show how topological photonic systems can protect the single or multi-photon states, which has a great potential for fault-tolerant on-chip quantum information processing and computing. We then discuss how quantum emitters could be interfaced with electromagnetic environments for creating topological quantum metamaterials with applications ranging from scalable quantum networks to strongly interacting topological phases of light.

## 4.1 Quantum information in topological metasurfaces

In the quantum regime, vacuum fluctuations of electromagnetic fields could be modified by the topological band structures of photonic systems. To show this, the authors in [286] considered a Hamiltonian with particle nonconserving terms that can coherently add and remove pairs of particles from the system and demonstrated the existence of a situation where the topologically protected chiral edge modes in a system with boundaries are unstable whereas all bulk modes are stable. Exploiting this exotic feature, the authors further showed that a topologically protected, quantum-limited traveling wave parametric amplifier could be realized. By manipulating the structure and vacuum fluctuations of topologically robust electromagnetic edge modes realized in a two-dimensional array of ring resonators under a uniform synthetic magnetic field emulating the integer quantum Hall effect, the authors in [251] experimentally realized a topological source of quantum correlated photon pairs via spontaneous four-wave mixing in which the spectral correlations between signal and idler photons are robust against fabrication disorder and outperforms a similarly designed topologically trivial source of correlated photons. In particular, to characterize the non-trivial correlations between the generated photons, the authors measured the second-order cross-correlation function $g_{s,i}^{(2)}(\tau)$ to detect signal and idler photons separated by time $\tau$ and observed a maximum $g_{s,i}^{(2)} \approx 80$ at $\tau = 0$. Furthermore, using a Hanbury-Brown-Twiss setup, the authors measured the conditional (heralded) autocorrelation function $g_{a,h}^{(2)}(\tau)$ for signal photons, conditioned on the detection of idler photons and observed a conditional $g_{a,h}^{(2)}(0) =$

0.20(8), indicating that the photons are anti-bunched (note that classical light sources are characterized by $g_a^{(2)}(0) \geq 1$ whereas quantum light sources, such as single photons, are distinguished by $g_a^{(2)}(0) < 1$). These interesting results open a route towards on-chip, scalable sources of heralded and entangled photons with identical spectra, which are promising for applications in quantum information processing and quantum communications.

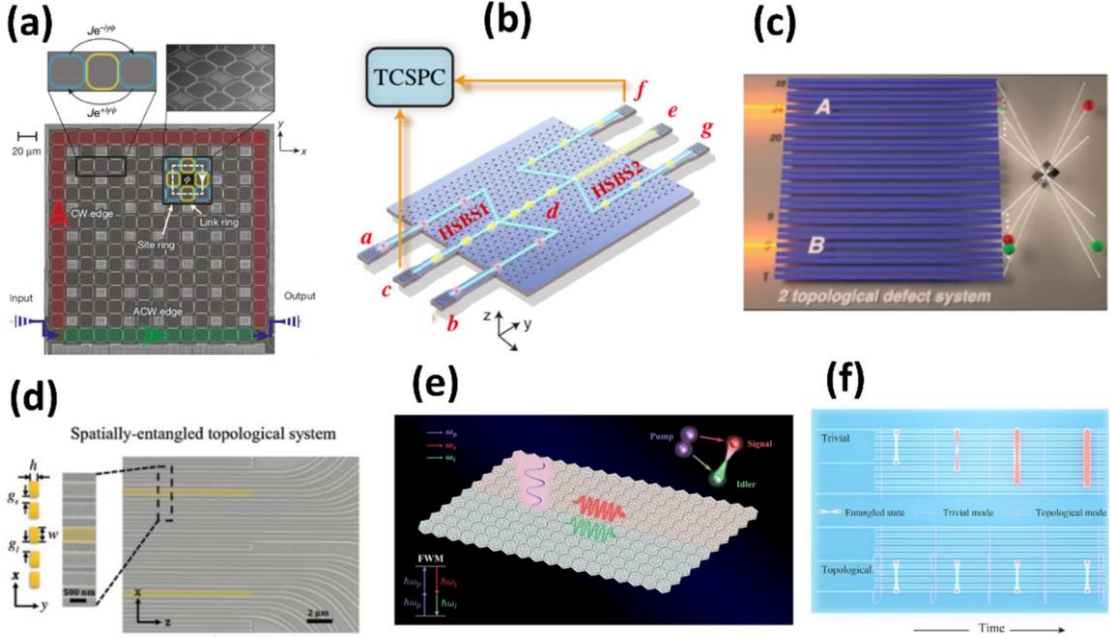

**Fig. 21.** Quantum information applications of topological photonics. (a) Topological source of quantum light in a coupled array of ring resonators [251]. (b) On-chip Hong-Ou-Mandel interference in a topologically protected valley-dependent quantum circuit [288]. (c) Topological protection of biphoton states in a nanophotonic platform [291]. (d) Topologically protected entangled photonic states in a nanophotonic platform with two topological defects [296]. (e) Topologically protected energy-time entangled biphoton states in spin-Hall topological photonic crystals [295]. (f) Topologically protected polarization quantum entanglement on a photonic chip [298].

Single photons emitted by quantum light sources are essentially indistinguishable, which can result in quantum interference. The authors later on [223] demonstrated a tunable source of indistinguishable photon pairs using dual-pump spontaneous four-wave mixing in a similar two-dimensional array of resonators emulating anomalous quantum Hall effect, i.e., in a configuration where while the magnetic flux through a single plaquette is non-zero, the average magnetic flux through a unit cell of two plaquettes of the lattice is zero. In this dual-pump spontaneous four-wave mixing process, two pump photons at different frequencies annihilate and create two frequency-degenerate photons, called the signal and the idler that are indistinguishable. Hong-Ou-Mandel (HOM) interference between the indistinguishable photon pairs was successfully demonstrated with a HOM dip of visibility reaching 88(10)%. A novel feature of the indistinguishable photon pairs generated in this setup is that their spectral-temporal bandwidth could be tuned by the input pump frequencies in the edge band and the generated photon pairs are energy-time entangled. Quantum interference of topological states of light has also been experimentally observed in integrated photonic circuits using coupled waveguide arrays [287] and valley-dependent quantum photonic circuits [288]. Especially, the authors in [287] implemented the off-diagonal Harper model using an array of coupled waveguides that exhibits topological boundary states.

Two single-photon topological boundary states, initially at opposite edges of a coupled waveguide array, are brought into proximity, where they interfere and undergo a beamsplitter operation. Hong-Ou-Mandel interference with 93.1 ± 2.8% visibility was observed in the experiments, confirming the non-classical behavior of the topological photonic states. Exploiting photonic valley states in a valley-dependent photonic platform, the authors in [288] demonstrated two-photon Hong-Ou-Mandel interference with a high visibility of 95.6 (0.6)%. The valley photonic platforms are CMOS compatible, scalable, and much more integrated, providing a novel method for on-chip valley-dependent quantum information process.

Topological photonic states can also protect the quantum features of photons, such as, single photon quantum state [289], squeezed light [290] and two-photon quantum correlations [291]-[293]. The authors in [289] demonstrated that the quantum features of single photons can be protected in topological structures against diffusion-induced decoherence in coupled waveguides and noise decoherence from the ambient environment. In specific, the authors considered a silica photonic chip of laser written waveguides implementing the "off-diagonal" version of Harper model and injected heralded single photons into the topological boundary states. To study whether the topological nontrivial boundary state can preserve the single-photon feature after the single photon outgoing from the lattices, the authors measured the second-order anticorrelation using Hanbury-Brown-Twiss interferometer and found that second-order anticorrelation parameter has no distinct changes before and after the photonic chip, confirming that the single-photon features can be well preserved.

Topological protection of squeezed light on a photonic chip was also demonstrated in [290], where the waveguide in the topologically protected lattices functions as a four-wave- mixing source in which glass absorbs two photons from the pump wave, and generates signal and idler photon pairs. To test whether the on-chip generated correlated photons can also be protected, the authors measured the cross-correlation function of the correlated photon pairs and found the topologically protected structure can provide nearly five times higher cross-correlation values than those in unprotected states. Furthermore, the squeezing parameters of the topologically protected channels are larger than that of the unprotected channels, indicating that the squeezed states can be well protected in the topological structure. Moreover, topology states can also protect biphoton states, e.g., the authors in [291] considered a topological lattice of silicon nanowire array on a silica substrate and demonstrated experimentally the robustness of the spatial features and the propagation constant of biphoton states. Exploiting a topological lattice fabricated in a borosilicate glass via the femtosecond laser direct writing technique, the authors in [292] experimentally demonstrated the topological protection of two-photon quantum states against decoherence, where by measuring the quantum correlation of two photons from the topologically nontrivial boundary state, a high cross-correlation and a strong violation of Cauchy-Schwarz inequality up to 30 standard deviations was obtained. The authors further showed that the topological protection is robust to the wavelength difference as well as distinguishability of two photons by preparing different quantum sources. Two-particle quantum correlations of indistinguishable photons were also studied experimentally in [293], where the authors found that the existence of the edge state at the topological edge in the SSH model has a significant influence on the quantum interference of indistinguishable photons.

Being the most counterintuitive quantum non-local correlations and an important resource of quantum information, computation and communications, quantum entanglement could also be protected by topology, which could be exploited for robust propagation of quantum information in disordered systems as demonstrated both theoretically [294][295] and experimentally [296]-[298] recently. In [294], the authors studied the transport of time-bin entangled photon pairs in coupled ring resonators simulating the integer quantum Hall effect and found that while the transport through edge states preserves temporal correlations of entangled photons, the transport through bulk does not preserve these correlations and can lead to significant unwanted temporal bunching or anti-bunching of photons. They further studied the transport in the presence of disorder and found that while the edge transport remains robust, enhanced bunching/anti-bunching of photons is a key feature for the bulk transport. Topological protection of entangled biphoton states has also been studied in valley photonic crystals [262], where the authors studied nonlinear four-wave mixing interaction of topological valley kink states propagating along the interface between two valley photonic crystals and demonstrated that the signal and idler photons generated from the four-wave mixing interaction are not only continuous frequency entangled, but also robust against the sharp bends and scattering, indicating topological protection of entangled photon pairs. Later on, the authors in [295] studied the quantum correlation between signal and idler photons generated from the FWM process in topological photonic crystals emulating the quantum spin Hall effect and found that the generated signal and idler photons are energy-time entangled, i.e., entangled between the frequencies of photon pair and their time of arrival. Furthermore, the topological edges states of the pump, signal and idler photons are robust against the sharp bends and defects, indicating that the energy-time entangled biphoton states are topologically protected.

Experimentally, topologically protected entangled photonic states have been studied in photonic chips with coupled waveguides [296]-[298]. The first experimental demonstration of topological protection of entangled biphoton states was reported in [296], where the authors considered a photonic chip of coupled silicon nanowires on a silica substrate and entangled biphoton states were generated via spontaneous four-wave mixing (SFWM) within the photonic chip itself. By studying the biphoton correlation map at the output of the topological lattices with different levels of deliberately introduced disorder, the authors demonstrated that the entanglement of the biphoton states can be well preserved against disorder that respects the chiral symmetry of the system. Biphoton entanglement of photonic modes of two different topologies was studied in [297], where the authors considered a bipartite array of silicon-on-insulator waveguides in which nonlinearly generated photon pairs are created in a superposition of three colocalized modes of two distinct topologies (both trivial and non-trivial). They also studied the average relative weight of the topological and trivial components of the entangled state when increasing levels of disorder and found that at low levels of disorder, the population probability of the trivial biphoton modes is higher than that of the topological mode whereas at strong disorder, the population probability of the topological biphoton mode is higher than that of any of the hybrid trivial modes. As a result, due to the coexistence of topological and trivial biphoton modes in such entanglement, the measured entangled biphoton state does not preserve its shape under the presence of off-diagonal disorder. Recently, the topological protection of two-photon polarization entangled states was studied in a photonic chip made of coupled waveguide array [298]. Employing the process tomography technique, the authors demonstrate that quantum entanglement can be well preserved by the topological states even when the chip material introduces disorder and relative polarization rotation in phase space.

## 4.2 Topological metasurface interfaced with quantum emitters
The interaction of quantum emitters with photonic environments plays an important role in quantum systems engineering, e.g., it could be used to build quantum metamaterials, where the fundamental building blocks of the artificial optical materials are quantum coherent entities and electromagnetic modes of the system and importantly, the quantum coherence of these artificially engineered materials can maintain a time longer than the traversal time of the relevant electromagnetic signal. Controllable interaction between quantum emitters is also an essential ingredient for the realization scalable quantum networks, where each quantum node could be a hybrid of photon and quantum emitter, as photon could be serving as flying qubit to transmit the information whereas quantum emitter could be used as local qubit to store the information. Moreover, due to the inherent nonlinearity of constituent quantum emitters, integrating quantum emitters into topological photonic structures could not only allow the study of light-matter interactions in topologically nontrivial environments but also provide an intriguing opportunity for exploring strongly correlated states of light. Thus, how quantum emitters will couple with the various topological photonic platforms has attracted great attention recently.

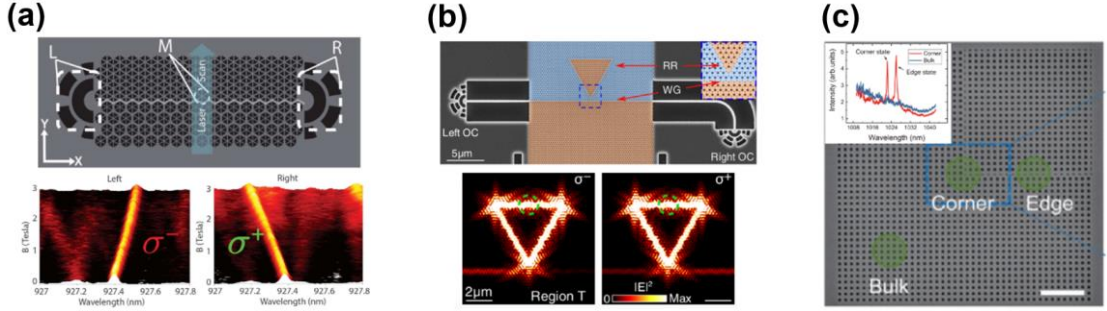

**Fig. 22.** Coupling of a single quantum emitter to topological photonic systems. (a) Chiral coupling between the helical topological edge modes of spin-Hall topological photonic crystals and a quantum emitter [46]. (b) Chiral topological photonics with an embedded quantum emitter in a valley-Hall topological photonic crystal waveguide [302]. (c) Coupling between a quantum emitter and second-order topological corner state for cavity quantum electrodynamics [305].

In [46], the authors experimentally demonstrated chiral coupling between a quantum emitter and the helical topological edge mode at the boundary of two distinct topological photonic crystals emulating the quantum spin Hall effect. As shown in Fig. 22(a), the authors considered a thin GaAs membrane with epitaxially grown InAs quantum dots at the center acting as quantum emitters. Under a magnetic field that induces a Zeeman splitting in the quantum dot excited state, two nondegenerate states that emit with opposite circular polarizations can couple to the helical topological edge modes and the emitted single photons exhibit robust transport, even in the presence of a bend.

Quantum emitter could also be used to probe the topological modes of photonic structures as demonstrated in [299], where the authors fabricated topological ring resonators in a spin Hall-type topological photonic crystal platform and by using spatially resolved PL measurements, topological modes confined to the PhC interface were visualized. However, as the band inversion point in spin-Hall topological waveguides happens around k=0, the topological edge modes in these slab platforms have radiative coupling to free-space modes. To avoid this radiative loss, valley-Hall topological photonic interface could be exploited as the topological valley edge modes lie below the light line. In [300], valley photonic crystal waveguide in a GaAs slab was experimentally characterized by the near-infrared light emitted from the InAs quantum dots inserted at the middle of the slab layer as an internal light source.

Robust wave propagation against sharp bends was further observed by the photoluminescence image from the quantum dots (QDs). Employing an interface between two valley-Hall topological photonic crystals, a topological resonator in the shape of a super-triangle as studied in [301], where chiral spontaneous emission with the direction of the emission depending on the polarization of the emitted light was demonstrated. More specifically, a quantum dots excited by external pump sources, is coupled with a topological resonator that supports two counterpropagating modes with opposite chiralities. By exciting the QDs and collecting the emission from the ends of the waveguide through the grating couplers, it was experimentally observed that the quantum emitter emits preferably into one of the counterpropagating edge modes depending upon the spin of the pump source. Collecting the photoluminescence (PL) signal at the left grating coupler, a single branch of the Zeeman split QD spectrum was clearly observed, which was a signature of chiral coupling. Furthermore, Purcell enhancement of emission due to the strong coupling between the quantum emitter and the topological resonator was also achieved when tuning the quantum emitter into resonance with the topological resonator by a magnetic field.

In another work [302], a chiral quantum optical interface using semiconductor QDs embedded in a valley-Hall topological PhC waveguide was also realized, where chiral coupling of single QDs to the nontrivial waveguide mode with a spin-dependent, averaged directional contrast of up to $0.75 \pm 0.02$ was demonstrated. As demonstrated in Fig. 22(b), chiral coupling of a QD with a photonic ring resonator in the form of a triangle with three 120-deg corners was also demonstrated, where the resulting experimental PL map was used to reveal the distinct triangular shape of the full waveguide. It also shows that the direction of emission into the waveguide relies on the dipole polarization, when a QD is located at the domain wall. Moreover, the measured photoluminescence signal verifies the chiral characteristics for the topologically protected ring resonator modes.

Most recently, an integrated topological add-drop filter consisting of a compact resonator coupled to a pair of access waveguides, was experimentally realized within a valley-Hall topological PhC [303]. The authors demonstrated chiral emission from a QD embedded within the resonator by showing that while one spin state of a QD transition couples into two of the four output ports of the device, the orthogonal spin state couples to the other two output ports. Slow light waveguides are preferable for high-performance single-photon sources as the slow light modes can enhance light-matter interactions and thus accelerate the emission of quantum emitters into the modes via the Purcell effect.

Recently, single-photon sources using single quantum dots (QDs) embedded in topological slow light waveguides based on valley photonic crystals were reported in [304], where the authors observed that single QDs coupled to the topological slow light waveguide exhibit large Purcell enhancement of spontaneous emission rate up to a factor of ~ 12 with a group index over 20. To confirm the single-photon nature of the QD emission, the authors further performed intensity autocorrelation measurements using a Hanbury-Brown-Twiss setup and clear antibunching behavior was observed. Apart from coupling to the topological photonic edge modes, quantum dots have also been coupled to second-order corner states as demonstrated in Fig. 22(c) [305], where enhancement of photoluminescence intensity by a factor of about 4 and emission rate by a factor of 1.3 are both observed when the quantum dot is on resonance with the corner state. Furthermore, the existence of topological corner and edge states with high Q was also observed by photoluminescence (PL) spectra.   The realization of coupled quantum emitter with topological corner state enables the application of topology into cavity quantum electrodynamics setup.

Apart from the coupling of a single quantum emitter to topological photonic modes, strong coupling and entanglement between emitters could also be achieved via coupling to the same topological photonic mode. Recent, the authors in [306] theoretically studied a photonic crystal platform supporting both topologically protected edge state and zero-dimensional topological corner cavities. When the quantum emitters are put into topological cavities, the strong coupling between them can be fulfilled with the assistance of the topologically protected edge state. Such a strong coupling could maintain a very long distance and be robust against various defects. Moreover, the duration of quantum beats for such entanglement can reach several orders longer than that for the entanglement in a conventional photonic cavity, making it beneficial for a scalable quantum-information process.

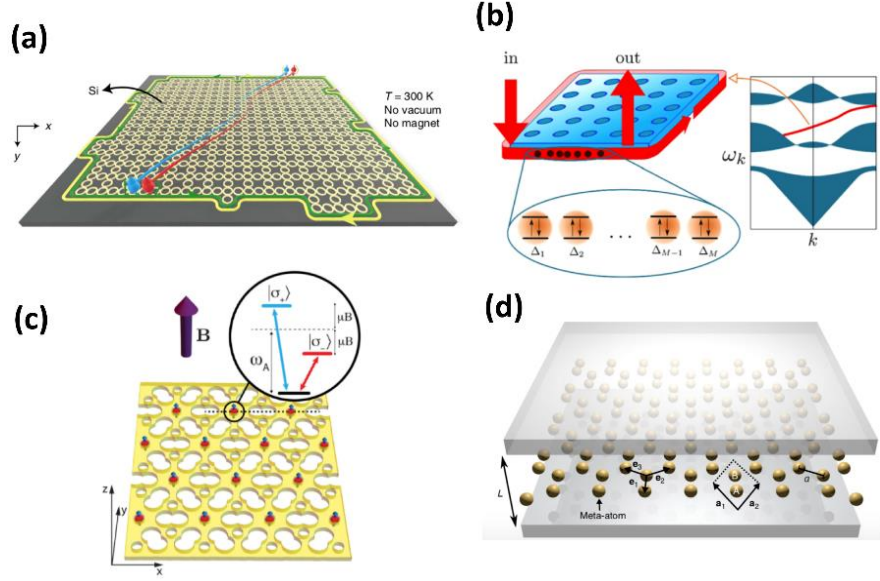

**Fig. 23.** Multiple quantum emitters in structured photonic environments for engineering of topological quantum metamaterials. (a) Topologically protected quantum entanglement emitters in a coupled array of ring resonators [108]. (b) 1D array of quantum emitters coupled to the chiral edge state of a topological photonic crystal [307]. (c) Topological quantum optics using 2D array of quantum emitters coupled to a photonic crystal slab [310]. (d) 2D array of quantum emitters embedded in a photonic cavity [314].

Most recently, topologically protected entanglement emitters were realized experimentally in a monolithically integrated plug-and-play silicon photonic chip in ambient conditions [108], as shown in Fig. 23(a). Exploiting the intrinsic four-wave mixing (FWM) nonlinearity of silicon waveguides, the system emits a topological Einstein-Podolsky-Rosen state, which was demonstrated to be topologically protected against artificial structure defects by comparing the state fidelities of $0.968 \pm 0.004$ and $0.951 \pm 0.010$ for perfect and defected emitters, respectively. The authors further verified the topological EPR entangled states by tomographic measurement and Bell violation, and verified the topological multiphoton NOON entangled states by interferometric measurement of the de Broglie wavelength.

Coupling a 1D array of quantum emitters to the chiral photonic edge state allows the creation of strongly correlated states of photons in a highly controllable way. In [307], the authors showed that photons in such chiral channel interacting with an ensemble of quantum emitters can produce outgoing states with robust, controllable, and universal properties originating from the topological nature of the edge state, as illustrated in Fig. 23(b). In particular, they found that the outgoing photonic wavefunction does not contain any information about the positions of emitters; its nodes are rather determined by universal numbers—the zeroes of Laguerre polynomials. In the case of two-photon scattering, they observed a clearly pronounced parity effect with respect to the number of emitters, which manifests in a transition from photons' bunching to antibunching as one changes the parity of the number of emitters from even to odd.

Further arranging the quantum emitters into regular 2D array can be used to build the quantum metasurface [308], in which quantum operator-valued reflectivity can be used to control both the spatiotemporal and quantum properties of transmitted and reflected light. In [309], the authors studied a two-dimensional honeycomb array of closely spaced quantum emitters and found that the system maintains topologically protected confined optical modes that are immune to large imperfections as well as to the most common loss processes such as scattering into free-space modes and spontaneous emission. Such modes can be used to control individual atom emission, and to create quantum nonlinearity at a single photon level, which could lead to strong interactions between individual excitations. However, to realize such topologically protected confined optical modes requires deeply subwavelength interatomic spacing, which may be challenging to implement experimentally.

Later on, the authors [310] considered triangular lattice of quantum emitter array integrated with a two-dimensional photonic crystal slab of air holes where the emitters can interact through the guided modes of the photonic crystal and found that in contrast to free-space realizations, very large topological band gaps and the existence of an almost completely flat topological band could be realized in this setup, in which the required lattice spacing of the emitters is comparable to the optical wavelength, as demonstrated in Fig. 23(c). However, the above works simplify the atom-photon interaction Hamiltonian by tracing out the photon degrees of freedom and applying the Markov approximation, which is not valid when the electromagnetic environment has narrow bandwidth features and is highly dispersive. The Markov approximation is also violated when the length of a spontaneously emitted photon becomes comparable to or smaller than the atom spacing, which might occur for stronger atom-photon coupling, resulting in faster decay rates. Moreover, another almost universal approximation in these models is the electric dipole approximation based on the assumption that the photo wavelength is much longer than the atomic size, which may not hold for surface plasmon polaritons with wavelengths far smaller than the free-space wavelengths. To overcome these difficulties and achieve unidirectional emission and nonreciprocal transmission of single photons, the authors in [311] studied the polaritonic band structure of two-dimensional atomic lattices coupled to a single excitation of a surface plasmon polariton mode using a method that does not suffer from the limitations of the Markov as well as dipole approximations and demonstrated that the setup allows the realization of topological gaps with arbitrary Chern numbers by manipulating the internal degree of freedom (angular momentum quantum number) of the atoms.

Quantum emitters can also interact with the bulk region of the photonic lattice rather than its edge and this was studied in [312], where the authors studied light-emitter interactions in two-dimensional photonic systems in the presence of a spatially homogeneous synthetic magnetic field for light and found for increasing magnetic field the photonic modes change from extended plane waves to circulating Landau levels, leading to the formation of strongly coupled Landau-photon polaritons. This phenomenon could be understood from the fact that an emitted photon cannot propagate away, but it is constrained to orbit around the emitter due to the effective Lorentz force. Moreover, by being composed of circulating and dispersionless, but still spatially extended photons, the spectral and dynamical features of these quasiparticles can be continuously tuned from a single-mode, cavity QED type behavior to that of a many-body system of strongly interacting particles in the presence of a magnetic field.

Topological photonic modes can also mediate exotic interaction between quantum emitters when they interact with such photonic baths, e.g., Dirac photonics has been the possibility of obtaining decoherence-free, power-law interactions between emitters when many of them couple to these types of structures. In [313], the authors developed a semianalytical theory of dipole radiation near photonic Dirac points in realistic structures which allows the authors to compute the effective photon mediated interactions along the whole unit cell. Using this theory, the authors found positions where the nature of the collective interactions changes from being coherent to dissipative ones, which might lead to strong super/subradiant effects. Finally, cavity-embedded metasurfaces [314]-[316] have also been proposed to tailor the topological properties of regular arrays of quantum emitters.

For example, in [314], the authors showed that rather than tuning the lattice structure of an emitter array, manipulating the surrounding photonic environment can also be used to tune the topological properties of the system. As shown in Fig. 23(d), the authors considered a honeycomb array of quantum emitters supporting two distinct species of massless Dirac polaritons, namely type-I and type-II and demonstrated that by modifying only the photonic environment via an enclosing cavity allows one to manipulate the location of the type-II Dirac points, leading to qualitatively different polariton phases. This novel feature enables one to alter the fundamental properties of the emergent Dirac polaritons while preserving the lattice structure—a unique scenario that has no analog in real or artificial graphene systems. Applying this method to a strained honeycomb metasurface that creates pseudo-magnetic fields [315], the authors demonstrated that the pseudo-magnetic field can be tuned by modifying the surrounding electromagnetic environment via enclosing the metasurface array of quantum emitters in a cavity waveguide, without altering the strain configuration. This method can also induce topological transition in arrays of quantum emitter coupled to a cavity waveguide as demonstrated in [316], where the authors considered a kagome metasurface of quantum emitters that is embedded in a cavity waveguide and showed that varying the cavity width modifies the nature of the dipole-dipole interactions which enables one to manipulate the Berry curvature and invert the valley-Chern numbers without inverting the symmetry-breaking perturbation, i.e., one can switch the chirality of the polariton valley-Hall edge states entirely by only changing the local photonic environment.

## 5. Conclusion and perspective

In this review, we have introduced the recent developments of topological metasurface from passive to active then to quantum. First, passive topological metasurfaces based on different physics and phenomena are surveyed. To realize the analog quantum Hall TM, an external magnetic field is needed to break the time-reversal symmetry. Time-reversal preserving TM without the need of magnetic fields could also be realized, e.g., the analog quantum spin Hall TM, quantum valley Hall TM, and Floquet TM. While the above phenomena are mostly based on the bulk-edge correspondence principle, other TM, such as based on BICs and Skyrmions have also been studied. Different from passive topological metasurfaces that mainly focus on a static or predefined functionality, active topological metasurfaces can have many advantages in practices and as examples, we have discussed nonlinear topological metasurfaces and reconfigurable topological metasurfaces. For nonlinear topological metasurfaces, we have discussed nonlinear frequency conversion as well as topological lasers. For reconfigurable topological metasurfaces, different ways to realize the reconfigurability, such as electrical, optical, mechanical, thermal, have been examined. Finally, we have discussed how topological metasurfaces can advance the study of quantum optics, e.g., TM can not only protect single or multi-photon states, but also integrate with quantum emitters to create strongly interacting topological quantum metasurfaces.

The research of topological metasurfaces is a fast-evolving area and before we end this review, we would like to highlight some current challenges and future directions. Regarding to the passive topological metasurfaces, the topological properties of the analog quantum Hall topological metasurfaces are the most robust ones due to the fact that applied external magnetic fields break the time-reversal symmetry and remove the backscattering channels completely. Currently, analog quantum Hall topological metasurfaces are most realized in the microwave regimes due to the weak magnetic-optical responses at higher frequencies. It remains to be seen whether materials with strong magnetic-optical responses could be synthesized at optical frequencies which would open a new chapter for possible applications of analog quantum Hall topological metasurfaces.

On the other hand, though analog quantum spin Hall topological metasurfaces and analog quantum valley Hall topological metasurfaces can be realized in all-dielectric materials without the magnetic fields, the robustness of the corresponding topological edge states is diminished. Moreover, the current studies of analog quantum valley Hall topological metasurfaces are mostly based on the hexagonal lattice symmetry where the Dirac cones and thus the valleys are pinned at K/K' points of the Brillouin zone. It would be interesting to explore the scenarios where the positions of the valleys could be tuned freely in the Brillouin zone and to study the robustness of the resulting valley edge states when the positions of the valleys are tuned.

Regarding to the Floquet topological metasurfaces, though both dynamic and effective modulations have been proposed to realize Floquet TM in the literature, however, up to now, these based on dynamic modulations have not been experimentally demonstrated, due to the extremely weak electro-optical and nonlinearity of materials, especially at high frequencies. On the other hand, the effective modulations have been experimentally demonstrated, however they are so far limited in coupled ring resonator scheme. In future, more efforts would be needed on advanced experimental technologies for dynamically-modulated TM and effective-modulation schemes beyond coupled ring-resonators array.

For high-order topological metasurfaces, currently, most of the topological corner states are realized via the nontrivial bulk dipole moment and the original pi-flux model for realizing topological corner states based on nontrivial bulk quadrupole moments is not very convenient for optical metasurfaces as one would need to find a way to implement the negative coupling in the pi-flux model. In future, proposing new models without negative coupling that host nontrivial bulk quadrupole moments and could be implemented using all-dielectric materials for optical metasurfaces, would be interesting for high-order topological metasurfaces.

For topological metasurfaces based on BICs and skyrmions, the super-high Q factors possessed by BICs can be used to remove radiation loss during light propagation, enhance nonlinear conversion efficiency, and improve the sensitivity of sensors. Meanwhile, BICs also introduce new possibilities for lasers and give rise to highly directional lasers, vortex lasers, and lasers exhibiting ultrafast all-optical switching. Optical skyrmion-related concepts could open a new avenue for topological photonics and can lead to new possibilities in light manipulations, e.g., polarization skyrmions offer the freedom to generate almost any polarization of light within a narrow physical space. Besides continuous-waves, skyrmion-related concepts can also be generalized to space-time wave packets, which can propagate nondispersive in free space or even dispersive media and thus play the role of information carriers in communications with light. Moreover, it is worth to notice that electrical circuit is another important platform to study exotic topological physics, leading to the emergence of topological circuits [317]-[322]. Due to high degrees of freedom and convenient experimental implementations, topological circuit is a promising platform to study active and nonlinear topological systems as well as high-dimensional and non-abelian topological insulators.

Active topological metasurface has promising applications in the development of advanced photoelectric devices. However, to further facilitate the applications of active topological metasurface, there are several challenges deserved to be addressed. In nonlinear topological metasurfaces, the high-power stimulation and low conversion efficiency are still two primary bottlenecks in the development of low-power and high-performance nonlinear devices in practical applications. In reconfigurable metasurfaces, it is very important to develop compact, intelligent, ultrafast controllable systems. For instance, recent studies have proposed some compact reconfigurable topological metasurface based on the complementary metal oxide semiconductor (CMOS) integrated technology. However, the integration of topological metasurface with its external control part is still a challenge. Moreover, we can now precisely control each unit cell of topological metasurface. But, as the increment of unit cells, it would be able and worthful to develop a smart approach to design intelligent topological metasurface for the implementation of more sophisticated functionalities.

Finally, harnessing the power of topological metasurfaces for future quantum technological applications is very promising as topological features of metasurfaces can effectively protect the fragility of quantum states. While TM coupled with a few quantum emitters have been demonstrated in experiments, the coupling of TM with a regular array of quantum emitters have not been demonstrated and the realization of such strongly interacting topological quantum metasurfaces would allow the exploration of many interesting problems from topological quantum computing to strongly interacting topological phases of light. We believe topological metasurfaces will continue to be an active and exciting area of research in the years to come both in terms of fundamental research and practical applications.

## Funding

## Data availability

No data were generated or analyzed in the presented research.

## Disclosures

The authors declare no conflicts of interest.